\documentclass[12pt]{article}
\usepackage{natbib,setspace,lscape,longtable}
\usepackage{graphicx,float}
\usepackage{amsmath,amsthm,amssymb,color}
\usepackage{subcaption}
\usepackage{geometry}
\usepackage[bottom]{footmisc}
\usepackage{indentfirst}
\usepackage{endnotes}
\usepackage{verbatim}
\usepackage{threeparttable}
\usepackage{comment}
\usepackage{algorithm}
\usepackage{titlesec}
\usepackage{algorithmic}
\usepackage{multirow}
\usepackage{makecell}

\usepackage{multibib}
\usepackage{natbib}
\usepackage{titling}
\usepackage{chngcntr}

\usepackage{dsfont}
\usepackage{hyperref}
\RequirePackage[mathlines]{lineno}
\usepackage{multirow}
\usepackage[titletoc]{appendix}
\bibpunct{(}{)}{;}{a}{,}{,}
\usepackage{array}
\newcolumntype{L}[1]{>{\raggedright\let\newline\\\arraybackslash\hspace{0pt}}m{#1}}
\newcolumntype{C}[1]{>{\centering\let\newline\\\arraybackslash\hspace{0pt}}m{#1}}
\newcolumntype{R}[1]{>{\raggedleft\let\newline\\\arraybackslash\hspace{0pt}}m{#1}}

\floatstyle{ruled}
\newfloat{algorithm}{tbhp}{loa}
\floatname{algorithm}{Algorithm}
\newcites{New}{References}

\newcommand{\bZ}{\mathrm{\bf Z}}

\newcommand{\bI}{\mathrm{\bf I}}

\newcommand{\bU}{\mathrm{\bf U}}
\newcommand{\bX}{\mathrm{\bf X}}
\newcommand{\bY}{\mathrm{\bf Y}}

\newcommand{\bzero}{\mathrm{\bf 0}}

\newcommand{\bmu}{\mbox{\boldmath $\mu$}}

\newcommand{\bgamma}{\mbox{\boldmath $\gamma$}}
\newcommand{\bGamma}{\mbox{\boldmath $\Gamma$}}

\newcommand{\bSigma}{\boldsymbol{\Sigma}}

\newcommand{\Cov}{\mathrm{Cov}}

\newcommand{\tr}{\mathrm{tr}}

\newcommand{\var}{\mathrm{var}}
\newcommand{\Var}{\mathrm{Var}}

\newcommand{\beq}{\begin{eqnarray*}}
\newcommand{\eeq}{\end{eqnarray*}}

\titleformat{\section}{\normalfont\Large\bfseries}{\thesection}{0.5em}{}
\titlespacing*{\section} {0pt}{5pt}{3pt}
\titlespacing*{\subsection} {0pt}{5pt}{2pt}
\numberwithin{equation}{section}
\theoremstyle{plain}

\newtheorem{prop}{Proposition}[section]

\theoremstyle{definition}

\newtheorem{remark}{Remark}[section]
\def\ben{\begin{equation*}}
\def\een{\end{equation*}}
\def\bea{\begin{eqnarray}}
\def\eea{\end{eqnarray}}
\def\bean{\begin{eqnarray*}}
\def\eean{\end{eqnarray*}}
\def\bep{\begin{prop}}
\def\eep{\end{prop}}
\def\bc{\begin{center}}
\def\ec{\end{center}}

\def \tr {\mbox{tr}}

\def\lowcomma{_{\textstyle,}}
\def\lowperiod{_{\textstyle.}}

\newtheorem{theorem}{\bf Theorem}

\newtheorem{assump}{\bf Assumption}

\allowdisplaybreaks[4]
\begin{document}

\title{Power-enhanced simultaneous test of high-dimensional mean vectors and covariance matrices with application to gene-set testing}
\author{
Xiufan Yu$^1$, Danning Li$^{2}$, Lingzhou Xue$^3$, and Runze Li$^3$ \\
$^1$University of Notre Dame, 
$^2$Northeast Normal University, and\\
$^3$Pennsylvania State University
}
\date{ First Version: September, 2020;\\
This Version: July, 2021.
}
\maketitle{}

\pagestyle{plain}

\begin{abstract}
Power-enhanced tests with high-dimensional data have received growing attention in theoretical and applied statistics in recent years. Existing tests possess their respective high-power regions, and we may lack prior knowledge about the alternatives when testing for a problem of interest in practice. There is a critical need of developing powerful testing procedures against more general alternatives. This paper studies the joint test of two-sample mean vectors and covariance matrices for high-dimensional data. 
We first expand the high-power region of high-dimensional mean tests or covariance tests to a wider alternative space and then combine their strengths together in the  simultaneous test. We develop a new power-enhanced simultaneous test that is powerful to detect differences in either mean vectors or covariance matrices under either sparse or dense alternatives.  We prove that the proposed  testing procedures align with the power enhancement principles introduced by \cite{fan2015power} and achieve the accurate asymptotic size and consistent asymptotic power. We demonstrate the finite-sample performance using simulation studies and a real application to find differentially expressed gene-sets in cancer studies. Our findings in the empirical study are supported by the biological literature.
\end{abstract}


\section{Introduction} \label{sec: introduction}

Inferences on the equality of two distributions is of significant interest in a wide range of real applications. Genetic studies use the differential gene expression analysis to understand how genes are related to diseases \citep{wang2015high}. Medical image analysis examines the differential structure of image to diagnose abnormal tissues \citep{ginestet2017hypothesis}. Pharmaceutical researchers rely on the analysis of comparative clinical trial outcomes for drug discovery and development \citep{cummings2019alzheimer}. 

To make inferences on the discrepancies between two distributions, we usually consider their mean vectors and covariance matrices that characterize  commonly used  distributions, for example, the elliptical distributions \citep{anderson2003introduction}. Over the past decade, there has been significant progress in testing the equality of two mean vectors \citep{chen2010two,wang2015high,wang2019combined,chen2019two} or covariance matrices \citep{li2012two,zhu2017testing,chen2019multi} under the high-dimensional setting. Yet few works are capable of examining both mean vectors and covariance matrices.

However, in practice, we often do not know whether the discrepancies reside in mean vectors or in covariance structure. It has been recognized that mean tests are powerful to detect the differences in mean vectors but cannot detect the different covariance structure. In contrast, covariance tests are powerful to identify the differences in covariance structure but are incompetent to distinguish the differential structure of two mean vectors. Thus, it is crucial to develop a new simultaneous testing procedure that is powerful to detect differences in either mean vectors or covariance matrices. 

Let $\bX$ and $\bY$ be two $p$-dimensional populations with mean vectors $(\bmu_1,\bmu_2)$  and covariance matrices $(\bSigma_1, \bSigma_2)$, respectively. We consider the simultaneous test on the equality of mean vectors and covariance matrices of the two populations, that is,
\vspace{-1ex}
\begin{equation}\label{eq: H0}
H_0:\bmu_1 = \bmu_2 \ \text{  and  } \ \bSigma_1 = \bSigma_2.
\vspace{-1ex}
\end{equation}
In real-world applications such as genetic studies, the sample size is often less than a hundred, but the number of features can be thousands or even larger \citep{clarke2008properties}. Throughout this paper, we assume that the dimension $p$ is much larger than the sample size $n_1$ or $n_2$. The challenge of high dimensionality leads to fundamental difficulties in understanding the asymptotic behavior of test statistics.

Two different classes of alternatives (i.e., dense alternatives and sparse alternatives) have been explored in the high-dimensional hypothesis testing. For dense alternatives, the parameter space of interest is defined using the squared $\ell_2$ norm, that is, $\|\bmu_1-\bmu_2\|^2$ and $\tr\{(\bSigma_1-\bSigma_2)^2\}$, and the distributions under $H_0$ and $H_1$ are hard to distinguish when the nonzero entries of $\bmu_1-\bmu_2$ and $\bSigma_1-\bSigma_2$ are of about the same size in the absolute value \citep{chen2010two,li2012two}. For sparse alternatives, the parameter space of interest is defined using the entry-wise maximum norm, that is, $\|\bmu_1-\bmu_2\|_{\max}= \max_{1\leq i\leq p} {(\mu_{1i}-\mu_{2i})^2}$ and $\|\bSigma_1-\bSigma_2\|_{\max}= \max_{1\leq i, j\leq p} (\sigma_{1,ij}-\sigma_{2,ij})^2$, and the distributions under $H_0$ and $H_1$ are hard to distinguish when there are only a few large nonzero entries of $\bmu_1-\bmu_2$ and $\bSigma_1-\bSigma_2$  \citep{arias2011global,cai2013two}. The mathematical definitions of dense alternatives and sparse alternatives will be presented in Section 3.

In the literature, there only exist a few works on jointly testing means and covariances. In the classical setting with a fixed dimension $p$, the likelihood ratio test (LRT) was extensively studied in the multivariate analysis \citep{anderson2003introduction} when the samples come from normal distributions. When $p$ diverges proportionally as the sample size tends to infinity such that $p/\min\{n_1, n_2\}\rightarrow c$ for some $0<c\leq 1$, \cite{jiang2013central} studied the modified LRTs under the normal assumption and derived central limit theorems. The normal assumption was recently relaxed by \cite{niu2019lr}. To allow $p$ to diverge at a comparable rate as the sample size tends to infinity, i.e., 
$0<c<\infty$, \cite{liu2017simultaneous} proposed a new approach by replacing the entropy loss with the quadratic loss for covariance matrix estimation. \cite{hyodo2018test} proposed a new joint test using a weighted sum of multiple U-statistics to allow $p$ to diverge  faster than the sample size.

However, most of these existing testing procedures only allow for a moderate-high dimension in the asymptotic regime such that the dimension diverges at a slower rate than the sample size. Also, these existing testing procedures are mainly based on the modified LRTs or the $L_2$-norm-based test. Like quadratic-form test statistics, they perform well against dense alternatives but perform poorly against sparse alternatives \citep{fan2015power,li2015joint,yu2020fisher}. These tests suffer from the power loss in detecting sparse signals, as the errors in estimating high-dimensional parameters accumulate in \citep{fan2015power}. Moreover, these joint testing procedures are essentially based on a weighted sum of one test statistic related to the mean difference and another test statistic related to the covariance difference. The weighted sum is not an ideal combination due to potentially different scales of two test statistics. These tests could be driven by the test statistic of a larger scale, leading to undesired power loss in the corresponding alternative space \citep{xie2011confidence}.

This paper aims to develop a new power-enhanced simultaneous testing procedure that is powerful to detect differences in either mean vectors or covariance structure against either sparse alternatives or dense alternatives under a high dimensional setting. \cite{fan2015power} introduced the power enhancement framework for high-dimensional hypothesis testing, which consists of the following  {\emph{power enhancement (PE)  principles}}: (a) no size distortion; (b) the power-enhance test is at least at powerful as the original test; (c) the power is substantially enhanced under a more general alternative. In this work, we interpret the more general alternatives from the following two perspectives: 

\begin{enumerate}
    \item[(i)] expanding the high-power region of mean tests or covariance tests to a wider alternative space respectively. We aim to develop the power-enhanced tests against the union of their corresponding dense and sparse alternatives.
    
    \item[(ii)] extending the test capability to alternative spaces with respect to both mean vectors and covariance matrices. We aim to combine both strengths of two power-enhanced tests and  develop a joint test that is capable of detecting the difference from either mean vectors or covariance matrices.
\end{enumerate}

To expand the high-power regions, we construct power-enhanced tests for mean vectors and covariance matrices separately. We revisit the test statistics of \cite{chen2010two} and \cite{li2012two} that are constructed based on the estimators of the squared Euclidean distance of two sample mean vectors and the squared Frobenius distance of two sample covariance matrices, respectively. It is known that they are powerful to detect dense signals but unable to detect sparse signals \citep{chen2019two}. We introduce their respective PE components to effectively enlarge the high-power regions to the union of sparse and  dense alternatives. We show that the proposed power-enhanced tests satisfy three desired PE principles. It is worth pointing out that we need new ideas to deal with a more challenging setting than that in \cite{fan2015power}. The mechanism of enhancing test power via PE components is to add a constructed component to an asymptotically pivotal statistic, so that the resultant testing power is strengthened upon the original test. The construction of PE components relies on a screening over the marginal test statistics. \cite{fan2015power} employs a quadratic-form OLS-based statistic, whose marginal distributions are asymptotically normal. However, \cite{chen2010two} and \cite{li2012two} use degenerate U-statistics, and the distributions of their marginal test statistics are no longer asymptotically normal but rather a $\chi^2$ distribution under the null hypothesis. The asymmetrically distributed marginal statistics require additional attention in 
the design of PE components. To the best of our knowledge, this is the first work that constructs PE components based on degenerate U-statistics.

After expanding the high-power regions, we aim to combine their strengths to develop the power-enhanced simultaneous test to further enhance the test capability for jointly testing mean vectors and covariance matrices. We prove the asymptotic independence of two PE test statistics and then aggregate information from the two aspects via the combination of their respective $p$-values using Fisher's method \citep{Fisher1925}.  We also show that the proposed power-enhanced simultaneous test satisfies three PE principles. It is important to note that, unlike \cite{fan2015power}, \cite{li2015joint}, and \cite{yu2020fisher}, we do not require the stringent normal assumption or independent assumption when deriving the asymptotic independence result. Compared with \cite{fan2015power} and \cite{li2015joint}, our proposed test is scale-invariant and computationally efficient.

We study the theoretical properties under an ultra-high dimensional setting where the dimension may grow at a nearly exponential rate of the sample size. Moreover, we conduct simulation studies to compare the proposed test's numerical performances against several benchmark tests under various alternatives. In a real application, we further demonstrate the power of the proposed test to find differentially expressed gene-sets using an acute lymphoblastic leukemia dataset.   Our findings are supported by the biological literature.

The rest of this paper is organized as follows. Section \ref{sec: Preliminaries} presents the preliminaries, and Sections \ref{subsec: powerEnhancement} and \ref{subsec: joint-test} include the complete methodological details.
Theoretical properties, including the power enhancement properties, the asymptotic size and power analysis as well as the asymptotic optimality, are also established in these two sections.  Section \ref{sec: simulation} conducts simulation studies to demonstrate the finite-sample properties under different alternative hypotheses. Section \ref{sec: realdata} presents an empirical study on identifying differentially expressed gene-sets among various types of cancers. Section \ref{sec: conclusion} includes a few concluding remarks. All technical details are presented in the supplementary note.

\section{Preliminaries}\label{sec: Preliminaries}

Let $\bX $ be a $p$-dimensional random vector with mean  $\bmu_1 = (\mu_{11}, \cdots, \mu_{1p})'$ and  covariance  $\bSigma_1 = \left(\sigma_{1,ij}\right)_{p \times p}$, and $\bY $ be a $p$-dimensional random vector with mean  $\bmu_2 = (\mu_{21}, \cdots, \mu_{2p})'$ and covariance  $\bSigma_2 = \left(\sigma_{2,ij}\right)_{p \times p}$. Suppose that $\{\bX_{1},\cdots,\bX_{n_1}\}$ are independently and identically distributed (i.i.d.)  copies of $\bX$, and $\{\bY_{1},\cdots,\bY_{n_2}\}$ are i.i.d. copies of $\bY$ that are independent of $\{\bX_1,\cdots,\bX_{n_1}\}$. Now, we consider the high-dimensional mean test
\begin{equation}\label{eq: H0-mean}
    H_{0m}:\bmu_1=\bmu_2,
\end{equation}
and the high-dimensional covariance test
\begin{equation}\label{eq: H0-cov}
    H_{0c}:\bSigma_1=\bSigma_2,
\end{equation}
respectively. \cite{chen2010two} proposed the following quadratic-form statistic $M_{n_1,n_2}$ for testing if the two high-dimensional populations share the same mean vector in \eqref{eq: H0-mean}:
\begin{equation}\label{stat: mean}
M_{n_1,n_2}=\frac{1}{n_1(n_1-1)}\sum_{u\neq v}^{n_1}\left(\bX_{u}'\bX_{v}\right) +\frac{1}{n_2(n_2-1)}\sum_{u\neq v}^{n_2}\left(\bY_{u}'\bY_{v}\right) -\frac{2}{n_1n_2}\sum_u^{n_1}\sum_v^{n_2}\left(\bX_{u}'\bY_{v}\right).
\end{equation}
To test the equality of two covariance matrices in \eqref{eq: H0-cov}, \cite{li2012two} constructed their test statistic  based on the squared Frobenius norm of $\bSigma_1 - \bSigma_2$. Since  $\| \bSigma_1 - \bSigma_2\|_F^2 = \tr\left((\bSigma_1-\bSigma_2)^2\right) = \tr(\bSigma_1^2) + \tr(\bSigma_2^2) - 2\tr(\bSigma_1\bSigma_2)$, they proposed a test statistic $T_{n_1,n_2}$ in the form of linear combination of unbiased estimators
for each term, that is,
\begin{equation}\label{stat: cov}
T_{n_1,n_2} =A_{n_1}+B_{n_2}-2C_{n_1, n_2}.
\end{equation}
where $A_{n_1}$, $B_{n_2}$ and $C_{n_1,n_2}$ are the unbiased estimators for $\tr(\bSigma_{1}^2)$, $\tr(\bSigma_{2}^2)$ and $\tr(\bSigma_{1}\bSigma_{2})$, respectively.

The following assumptions are discussed in \citeNew{chen2010two} and \citeNew{li2012two} to establish the asymptotic properties of two test statistics $M_{n_1,n_2}$ and $T_{n_1,n_2}$.

\begin{assump}\label{assum: A2} 
For any $i,j,k,l\in \{1,2\}$, as $n_1,n_2,p\rightarrow\infty$,
  \begin{equation}\label{eq: assumA2}
  \tr\left( \bSigma_k\bSigma_l\right)\rightarrow\infty, \quad 
  \tr\left\{\bSigma_i\bSigma_j\bSigma_k\bSigma_l\right\}=o\left\{ \tr\left(\bSigma_i\bSigma_j\right)\tr\left(\bSigma_k\bSigma_l\right)\right\}\lowperiod
  \end{equation}
\end{assump}

\begin{assump}\label{assum: A3} 
The random vectors $\{\bX_u\}_{u=1}^{n_1}$, $\{\bY_v\}_{v=1}^{n_2}$ satisfy 
\begin{equation}\label{eq: assumA3-model}
    \bX_{u}=\bGamma_1 \bZ_{1u}+\bmu_1,\ \bY_{v}=\bGamma_2 \bZ_{2v}+\bmu_2\quad 1\leq u \leq n_1, 1\leq v \leq n_2,
\end{equation}
where $\bGamma_i = (\bgamma_{i1}, \cdots, \bgamma_{ip})'$ is a $p\times m_i$ matrix for some $m_i\geq p$ such that $\bGamma_i\bGamma_i'=\bSigma_i$ for $i=1,2$, and
$\{\bZ_{ij}\}_{j=1}^{n_i} = \{(z_{ij1},\cdots, z_{ijm_i})'\  \}_{j=1}^{n_i} \in \mathbb{R}^{m_i}$ are i.i.d. random vectors such that for any positive integers $q$ and $\alpha_l$ such that $\sum_{l=1}^q \alpha_l \leq 8$, and for any $1\leq k_1 \neq k_2 \neq \cdots \neq k_q \leq m_i$,
\begin{equation}\label{eq: assumA3-moments}
   E(z_{ijk}) = 0,\ \Var(z_{ijk})=1,\ \Cov(z_{ijk_1}, z_{ijk_2}) = 0,\  E(z^4_{ijk}) = 3 + \Delta_i,\  E(z_{ijk}^8) < \infty,
\end{equation}
and 
\begin{equation} \label{eq: assumA3-pseudo-indep}
    E(z_{ijk_1}^{\alpha_1}z_{ijk_2}^{\alpha_2}\cdots z_{ijk_q}^{\alpha_q}) = E(z_{ijk_1}^{\alpha_1}) E(z_{ijk_2}^{\alpha_2})\cdots E(z_{ijk_q}^{\alpha_q}).
\end{equation}
\end{assump}

Note that (\ref{eq: assumA3-model}) expresses the samples using a factor-model structure, and (\ref{eq: assumA3-moments}) spells the moment condition needed for the factors $z_{ijk}$, in which the $\Delta_i$ measures the fourth-moment difference compared to a standard normal distribution. (\ref{eq: assumA3-pseudo-indep}) depicts a pseudo-independence pattern among its components for each $\bZ_{ij}$. The condition is satisfied if $\bZ_{ij}$ does have independent structure.

Under the null hypothesis $H_{0m}$, \cite{chen2010two} considered the standardized the test statistic $M_{n_1,n_2}/\widehat{\sigma}_{01} $ and proved that,
\begin{equation}\label{eq: mean-stat-asymp-distribution}
    \text{under }H_{0m}: \quad \frac{M_{n_1,n_2}}{\widehat{\sigma}_{01}} \overset{d}{\rightarrow} N(0,1)\quad \text{as } n_1,n_2,p\rightarrow\infty,
\end{equation}
where $\widehat{\sigma}_{01}$ is a consistent estimator of  $\sigma_{01} = (\frac{2}{n_1(n_1-1)}\tr(\bSigma_1^2)+\frac{2}{n_2(n_2-1)}\tr(\bSigma_2^2)+\frac{4}{n_1n_2}\tr(\bSigma_1\bSigma_2))^{\frac{1}{2}}$, which is the standard deviation of $M_{n_1,n_2}$ under $H_{0m}$. The test rejects $H_{0m}$ with significance level $\alpha$ if 
    $M_{n_1,n_2} \geq \widehat{\sigma}_{01} z_{\alpha},$
where $z_{\alpha}$ is the upper $\alpha$-quantile of standard normal distribution.

Under the null hypothesis $H_{0c}$, we note that the leading variance of $T_{n_1,n_2}$ is 
    $\sigma_{02}^2 = 4\left(\frac{1}{n_1}+\frac{1}{n_2}\right)^2 \tr^2\left(\bSigma^2\right).$
With $\widehat{\sigma}_{02}$ being a consistent estimator of $\sigma_{02}$, \citeNew{li2012two} conducted the test for $H_{0c}$ on the basis of the test statistic $T_{n_1,n_2}/\widehat{\sigma}_{02} $ and proved that,
\begin{equation}\label{eq: cov-stat-asymp-distribution}
    \text{under }H_{0c}: \quad \frac{T_{n_1,n_2}}{\widehat{\sigma}_{02}} \overset{d}{\rightarrow} N(0,1)\quad \text{as } n_1,n_2,p\rightarrow\infty,
\end{equation}
The test rejects $H_{0c}$ with a nominal significance level $\alpha$ if 
    $T_{n_1,n_2} \geq \widehat{\sigma}_{02} z_\alpha.$

In the sequel, we will present our proposed power-enhanced simultaneous test on jointly testing means and covariances in high dimensions. In Section \ref{subsec: powerEnhancement}, we propose power-enhanced tests for the mean test and the covariance test respectively to boost their respective power. In Section \ref{subsec: joint-test}, anchored in these two power-enhanced statistics, we study their asymptotic joint distribution and subsequently introduce our simultaneous test to expand the test capability for jointly testing high-dimensional mean vectors and covariance matrices.

\section{Power-Enhanced Tests}\label{subsec: powerEnhancement}

Both $M_{n_1, n_2}$ and $T_{n_1, n_2}$ are quadratic-form statistics. It has been known that such type of statistics suffer from low power against sparse alternatives where the parameter of interest differs only in a small proportion of coordinates. One predominant approach to achieve high testing power against sparse alternatives is to utilize extreme values to construct test statistics \citep{cai2013two,chernozhukov2019inference}, whereas another way continues with the quadratic-form statistics but rules out non-signal bearing dimensions via thresholding \citep{fan1996test,chen2019two,chen2019multi}. However, these tests generally require either stringent conditions or bootstrap to derive the limiting null distribution and are likely to suffer from size distortions due to slow convergence. Also, even though the extreme value tests and thresholding tests retain high power against sparse alternatives, they tend to lack the ability to detect dense and faint signals, in which circumstances the quadratic-form tests are favored.

To deal with the challenge mentioned above, we first explore power enhancement for testing high-dimensional mean vectors and covariance matrices based on $M_{n_1 n_2}$ and $T_{n_1, n_2}$, respectively. \cite{fan2015power} provides a helpful insight for us to enhance testing power against sparse alternatives and preserve the merits of existing quadratic-form tests at the same time. We construct two PE components $J_m$ and $J_c$, which are designed to take zero values under the null hypothesis but diverge quickly under sparse alternatives. The PE components are designed delicately 
following the guidance of the three PE principles. By adding the PE components to the original statistics, the resultant tests $\widehat{\sigma}_{01}^{-1} M_{n_1,n_2} + J_m$ and $\widehat{\sigma}_{02}^{-1} T_{n_1,n_2} + J_c$ acquire substantially enhanced power under sparse  alternatives with little size distortion under the null hypothesis.

Different from \cite{fan2015power}, the distributions of our marginal test statistics are no longer asymptotically normal under $H_0$. To be more specific, \cite{fan2015power} uses an quadratic-form OLS-based statistic to test the significance of the intercept in multi-factor pricing models. For each coordinate, the marginal test statistic asymptotically follow a standard normal distribution. Yet here, $M_{n_1,n_2}$ and  $T_{n_1,n_2}$ are degenerate U-statistics. Under the null hypothesis, their marginal statistics are no long asymptotically normal, causing difficulties in designing the PE components. In specific, the PE component is usually constructed using a screening technique. A properly chosen threshold is critical to capture the signal-bearing dimensions while exclude non-signal-bearning dimensions effected by estimation noise. The choice of such threshold is straightforward for the well-know normal distribution but requires additional efforts for non-normal distributions. 
After careful investigation, we prove that the marginal standardized statistics follow chi-squared distributions. To overcome the challenge brought by these asymmetrically distributed marginal statistics, we control the tail probabilities using a generalized result \citep{petrov1954generalization} of Cram{\'e}r's limiting theorem, and choose the thresholds accordingly. 

Let $n = n_1 + n_2$ and  $\delta_{p}$ and $\eta_{p}$ be the thresholds chosen 
for the mean statistics and covariances statistics, respectively. We choose $J_m$ and $J_c$ to be the sum of marginal standardized statistics whose values exceed $\delta_{p}$ and $\eta_{p}$. By construction, the screening procedure rules out all the noises under the null hypothesis. Still, it makes it capable of capturing non-zero signals under sparse alternatives, implying that $J_m$ and $J_c$ equal to zero under the null hypothesis but diverge quickly under the sparse alternatives.

\subsection{Power-Enhanced Mean Tests}
We use $\bX = (X_1, \cdots, X_p)'$ and $\bY = (Y_1, \cdots, Y_p)'$ to denote the random vectors of interest. Let $\bX_{u} = (X_{u1}, \cdots, X_{up})'$ and $\bY_{v} = (Y_{v1}, \cdots, Y_{vp})'$ be the corresponding random samples. We rewrite the statistic $M_{n_1,n_2}$ into $M_{n_1,n_2}= \sum_{i=1}^p M_{i}$, 
where
\begin{equation*}
M_{i} = \frac{1}{n_1(n_1-1)}\sum_{u\neq v}^{n_1} \left(X_{ui}X_{vi}\right) +\frac{1}{n_2(n_2-1)}\sum_{u\neq v}^{n_2}\left(Y_{ui}Y_{vi}\right) -\frac{2}{n_1n_2}\sum_u^{n_1}\sum_v^{n_2}\left(X_{ui}Y_{vi}\right).
\end{equation*}
For each $i = 1,\cdots, p$, $M_i$ consistently estimates $(\mu_{1i} - \mu_{2i})^2$ as $n_1, n_2 \rightarrow \infty$. Under the null hypothesis $H_{0m}: \bmu_{1} = \bmu_{2}$, the variance of $M_i$ is
$$\nu_i := \frac{2}{n_1(n_1-1)}\sigma^2_{1,ii} + \frac{2}{n_2(n_2-1)}\sigma^2_{2,ii} + \frac{4}{n_1n_2}\sigma_{1,ii}\sigma_{2,ii}, $$
which can be consistently estimated by 
$ \widehat\nu_i := \frac{2}{n_1(n_1-1)}\widehat\sigma^2_{1,ii} + \frac{2}{n_2(n_2-1)}\widehat\sigma^2_{2,ii} + \frac{4}{n_1n_2}\widehat\sigma_{1,ii}\widehat\sigma_{2,ii},$
with $\widehat\sigma_{1,ii}$ and $\widehat\sigma_{2,ii}$ being sample variances of $X_i$ and $Y_i$, respectively.
Define
\begin{equation}
J_m = \sqrt{p}\sum_{i=1}^p M_i\widehat\nu^{-1/2}_i \mathcal{I}\{ \sqrt{2}M_i\widehat\nu^{-1/2}_i + 1 >  \delta_{p} \}
\end{equation}
with $\delta_{p} = 2\log p$ as the power enhancement component for the mean test. The theoretical analysis regarding $J_m$ is established upon $\delta_{p} = 2\log p$. In practical implementations, we follow \cite{fan2015power} to choose a slightly larger thresholding value, specifically $\delta_{p,n} =2\log p\log\log n$, to mitigate finite-sample biases.

In what follows, we present some theoretical properties of the constructed PE component $J_m$ as well as the proposed power-enhanced mean test. To ensure that adding the PE component does not bring in size distortion, \cite{fan2015power} assumes the errors in a regression model follow a normal distribution. Benefiting from the usage of concentration inequalities to analyze the tail probabilities for degenerate U-statistics, we only assume the distributions of both populations are sub-Gaussian. 
\begin{assump}\label{assum: sub-Gaussian} 
There exists a positive constant $H$ such that for all $h\in[-H, H]$,
\begin{equation}\label{eq: sub-Gaussian} 
 Ee^{h(X_{ui} - \mu_{1i})^2} < \infty, \ Ee^{h(Y_{vi} - \mu_{2i})^2} < \infty\quad \text{ for } i=1, \cdots, p.   
\end{equation}
\end{assump}
The sub-Gaussianity assumption is imposed to control the tail probability of marginal statistics, ensuring the PE components equal to zero under the null hypothesis. 
This condition has also been assumed in relevant literature such as \cite{chen2019multi}, \cite{chen2019two}.

\begin{theorem} \label{thm: PE-Mean}
Suppose $n_1/\left(n_1+n_2\right)\rightarrow \gamma$ for some constant $\gamma\in (0,1)$ as $\min\{n_1,n_2\}\rightarrow\infty$ and $\log p = o(n^{1/3})$. Given Assumptions \ref{assum: A2}-\ref{assum: sub-Gaussian},  under the null hypothesis $H_{0m}: \bmu_1 = \bmu_2$, as $n_1, n_2, p \rightarrow \infty$, 
\begin{equation}\label{eq: MPE-stat}
    P\left( J_m = 0 | H_{0m} \right ) \rightarrow 1, \quad  M_{PE} = \frac{1}{\widehat{\sigma}_{01}}\sum_{i=1}^p M_i + J_m \overset{d}{\rightarrow} N(0,1).
\end{equation}
\end{theorem}
Theorem \ref{thm: PE-Mean} proves that $J_m = 0$ holds under $H_{0m}$ with probability tending to 1. Thus, adding $J_m$ to the mean statistic $\widehat{\sigma}_{01}^{-1}M_{n_1, n_2}$ will not affect its limiting null distribution. The proposed power-enhanced mean test rejects $H_{0m}$ with the significance level $\alpha$ if $M_{PE} \geq z_{\alpha}$.

\subsection{Power-Enhanced Covariance Tests}
As for the covariance test statistic $T_{n_1, n_2}$, 
we first decompose $T_{n_1,n_2}$ into
\begin{equation*}
T_{n_1,n_2} =\sum_{i=1}^p \sum_{j=1}^p T_{ij}=\sum_{i=1}^p \sum_{j=1}^p ( A_{ij} + B_{ij} - 2C_{ij}),
\end{equation*}
where
\begin{align*}
\begin{aligned}
A_{ij} & = \frac{1}{n_1(n_1-1)} \sum_{u \neq v}^{n_1} X_{ui}X_{vi}X_{uj}X_{vj} - \frac{2}{n_1(n_1-1)(n_1-2)}\sum_{u \neq v \neq k}^{n_1} X_{ui}X_{vi}X_{vj}X_{kj} \\
& + \frac{1}{n_1(n_1-1)(n_1-2)(n_1-3)} \sum_{u \neq v \neq k \neq l}^{n_1} X_{ui}X_{vi}X_{kj}X_{lj}, \\
B_{ij} & = \frac{1}{n_2(n_2-1)} \sum_{u \neq v}^{n_2} Y_{ui}Y_{vi}Y_{uj}Y_{vj} - \frac{2}{n_2(n_2-1)(n_2-2)}\sum_{u \neq v \neq k}^{n_2} Y_{ui}Y_{vi}Y_{vj}Y_{kj} \\
& + \frac{1}{n_2(n_2-1)(n_2-2)(n_2-3)} \sum_{u \neq v \neq k \neq l}^{n_2} Y_{ui}Y_{vi}Y_{kj}Y_{lj}, \\
C_{ij} & = \frac{1}{n_1n_2}\sum_{u}^{n_1}\sum_{v}^{n_2} X_{ui}Y_{vi}X_{uj}Y_{vj} - \frac{1}{n_1n_2(n_1-1)}\sum_{u \neq k}^{n_1}\sum_{v}^{n_2} X_{ui}Y_{vi}Y_{vj}X_{kj} \\
& - \frac{1}{n_1n_2(n_2-1)}\sum_{u\neq k}^{n_2} \sum_v^{n_1} Y_{ui}X_{vi}X_{vj}Y_{kj} + \frac{1}{n_1n_2(n_1-1)(n_2-1)} \sum_{u \neq k}^{n_1} \sum_{v \neq l}^{n_2} X_{ui}Y_{vi}X_{kj}Y_{lj}.
\end{aligned}
\end{align*}

The decomposition is essential to derive the power enhancement. For each $i, j = 1,\cdots, p$, $T_{ij}$ consistently estimates the element-wise difference in covariances, i.e., $T_{ij} \overset{p}{\rightarrow} (\sigma_{1,ij} - \sigma_{2,ij})^2$ as $n_1, n_2 \rightarrow \infty $. Under the null hypothesis $H_{0c}: \bSigma_{1} = \bSigma_{2}$, the variance of $T_{ij}$ is
\begin{align*}
\xi_{ij} := &\  2\left(\frac{1}{n_1} \left( \sigma^2_{1,ij} + \sigma_{1,ii}\sigma_{1,jj} + \Delta_1\tr(\bgamma_{1i}\bgamma_{1j}^T \circ \bgamma_{1i}\bgamma_{1j}^T ) \right) \right. \\
&\qquad + \left. \frac{1}{n_2} \left( \sigma^2_{2,ij} + \sigma_{2,ii}\sigma_{2,jj} + \Delta_2\tr(\bgamma_{2i}\bgamma_{2j}^T \circ \bgamma_{2i}\bgamma_{2j}^T) \right) \right)^2 (1+o(1)),
\end{align*}
{
where $\bgamma_{ki}$ and $\Delta_k$, $k=1,2$ are defined by (\ref{eq: assumA3-model}) and (\ref{eq: assumA3-moments}) in Assumption \ref{assum: A3}. In addition, we know that $\var((X_i-\mu_{1i})(X_j-\mu_{1j})) = \sigma^2_{1,ij} + \sigma_{1,ii}\sigma_{1,jj} + \Delta_1\tr(\bgamma_{1i}\bgamma_{1j}^T \circ \bgamma_{1i}\bgamma_{1j}^T )$, and analogously, $\var((Y_i-\mu_{2i})(Y_j-\mu_{2j})) =\sigma^2_{2,ij} + \sigma_{2,ii}\sigma_{2,jj} + \Delta_2\tr(\bgamma_{2i}\bgamma_{2j}^T \circ \bgamma_{2i}\bgamma_{2j}^T) $. Therefore, $\xi_{ij}$ can be consistently estimated by 
}
{
\small
\begin{equation*}
\widehat\xi_{ij} := 2\left( \frac{1}{n^2_1} \sum_{u=1}^{n_1} \{(X_{ui}-\bar{X_i})( X_{uj}-\bar{X_j}) - \widehat\sigma_{1,ij}\}^2  + \frac{1}{n^2_2}  \sum_{v=1}^{n_2} \{(Y_{vi}-\bar{Y_i})( Y_{vj}-\bar{Y_j}) - \widehat\sigma_{2,ij}\}^2  \right)^2 \lowcomma
\end{equation*}
}
where $\bar{X_j}$ and $\bar{Y_j}$ are the sample mean of $X_j$ and $Y_j$, $\widehat\sigma_{1,ij}$ and $\widehat\sigma_{2,ij}$ are sample covariances of $(X_i, X_j)$ and $(Y_i, Y_j)$, respectively.
Define
\begin{equation}
J_c = \sqrt{p}\sum_{i=1}^p\sum_{j=1}^p T_{ij}\widehat\xi^{-1/2}_{ij} \mathcal{I}\{ \sqrt{2}T_{ij}\widehat\xi^{-1/2}_{ij} +1 >  \eta_{p} \}
\end{equation}
as the power enhancement component for the covariance test, with $\eta_{p} = 4\log p$.
Similar to the previous subsection, the theoretical analysis regarding $J_c$ is established upon $\eta_{p} = 4\log p$. In practical implementations, we use a slightly larger thresholding value, specifically $\eta_{p,n} =4\log p\log\log n$, for the purpose of mitigating finite-sample biases.

\begin{theorem} \label{thm: PE-Cov}Suppose $n_1/\left(n_1+n_2\right)\rightarrow \gamma$ for some constant $\gamma\in (0,1)$ as $\min\{n_1,n_2\}\rightarrow\infty$ and $\log p = o(n^{1/5})$. Given Assumptions \ref{assum: A2}-\ref{assum: sub-Gaussian}, under the null hypothesis $H_{0c}: \bSigma_1 = \bSigma_2$, as $n_1, n_2, p \rightarrow \infty$,
\begin{equation}\label{eq: TPE-stat}
    P\left( J_c = 0 | H_{0c} \right) \rightarrow 1, \quad T_{PE} = \frac{1}{\widehat{\sigma}_{02}}\sum_{i=1}^p\sum_{j=1}^p T_{ij} + J_c \overset{d}{\rightarrow} N(0,1).
\end{equation}
\end{theorem}
Theorem \ref{thm: PE-Cov} proves that under the null hypothesis $H_{0c}$, $J_c = 0$ with probability approaching 1. The power-enhanced covariance test rejects $H_{0m}$ with significance level $\alpha$ if $T_{PE} \geq z_{\alpha}$.

\subsection{Power Enhancement Properties}
In this subsection, we study the power enhancement properties of our proposed power-enhanced tests $M_{PE}$ and $T_{PE}$. \cite{chen2010two} and \cite{li2012two} provided power analysis of the mean test statistic  $M_{n_1,n_2}$ and the covariance test statistic  $T_{n_1,n_2}$, respectively. Consider the following parameter spaces  $\mathcal{G}_m^d$ and $\mathcal{G}_c^d$ for their alternative hypotheses:
\begin{align*}
\mathcal{G}_m^d & = \bigl\{(\bmu_1,\bmu_2):  \min\{n_1,n_2\}\|\bmu_1-\bmu_2\|^2/\sqrt{\max\{\tr(\bSigma_1^2),\tr(\bSigma_2^2)\}} \rightarrow \infty \bigr\}\lowcomma \\ 
\mathcal{G}_c^d & = \bigl\{ (\bSigma_1,\bSigma_2): \bSigma_1 >0, \bSigma_2>0, \frac{1}{n_1}\tr(\bSigma_1^2) + \frac{1}{n_2}\tr(\bSigma_2^2) = o\left(\tr\{(\bSigma_1-\bSigma_2)^2\}\right)\bigr\}\lowperiod
\end{align*}

\cite{chen2010two} pointed out that as $n_1,n_2,p\rightarrow \infty$, the mean test statistic $M_{n_1,n_2}$ would correctly reject the null hypothesis $H_{0m}$ with probability approaching 1 if the mean differences $\bmu_1-\bmu_2$ fall into the subspace $\mathcal{G}_m^d$. \cite{li2012two} drew analogous conclusions with regards to the covariance alternative space $\mathcal{G}_c^d$ corresponding to the covariance test $T_{n_1,n_2}$. More specifically, as $n_1,n_2,p\rightarrow \infty$,
\begin{equation}\label{eq: sep-power}
    \inf_{(\bmu_1, \bmu_2)\in \mathcal{G}_m^d} P\left(M_{n_1,n_2} \geq \widehat{\sigma}_{01}z_\alpha\right) \rightarrow 1\  \text{ and } \  \inf_{(\bSigma_1,\bSigma_2)\in \mathcal{G}_c^d } P\left(T_{n_1,n_2} \geq \widehat{\sigma}_{02} z_\alpha\right) \rightarrow 1. 
\end{equation}

Note that $\mathcal{G}_m^d$ and $\mathcal{G}_c^d$ use the squared Euclidean-norm $\|\bmu_1-\bmu_2\|^2$ and the squared Frobenius-norm $\|\bSigma_{1}-\bSigma_{2}\|_F^2$ to specify a large magnitude of differences in mean vectors and covariance matrices in order for the tests to be powerful in detecting the discrepancies. 

In what follows, we present the power enhancement properties of our proposed tests. We will show that adding the power enhancement components $J_m$ and $J_c$ enables the tests to observe sparse signals which only differ in a few coordinates.

\begin{theorem} \label{thm: PE}
Suppose $n_1/\left(n_1+n_2\right)\rightarrow \gamma$ for some constant $\gamma\in (0,1)$ as $\min\{n_1,n_2\}\rightarrow\infty$ and $\log p = o(n^{1/5})$. Given Assumptions \ref{assum: A2}-\ref{assum: sub-Gaussian}, as $n_1, n_2, p \rightarrow \infty$, we have
$$\inf_{(\bmu_1, \bmu_2)\in \mathcal{G}_m^d \cup \mathcal{G}_m^s} P\left(M_{PE} \geq z_\alpha\right) \rightarrow 1, \quad \text{and} \quad \inf_{(\bSigma_1, \bSigma_2)\in \mathcal{G}_c^d \cup \mathcal{G}_c^s} P\left(T_{PE} \geq z_\alpha\right) \rightarrow 1,$$
with 
\begin{align*}
\mathcal{G}_m^s = &\bigl\{(\bmu_1,\bmu_2):  \max_{1\leq i\leq p} \frac{(\mu_{1i}-\mu_{2i})^2}{\nu_i^{1/2}} \geq C \delta_{p} \bigr\} \\
\mathcal{G}_c^s = &\bigl\{ (\bSigma_1,\bSigma_2): \bSigma_1 >0, \bSigma_2>0, \max_{1\leq i, j\leq p} \frac{(\sigma_{1,ij}-\sigma_{2,ij})^2}{\xi_{ij}^{1/2}} \geq C \eta_{p} \bigr\}
\end{align*}
where $C$ is an absolute constant that does not depend on $n_1$, $n_2$ and $p$. 
\end{theorem}

Theorem \ref{thm: PE} shows that the power-enhanced tests have the same rejection regions as those of the original tests, but the high power regions are substantially expanded from $\mathcal{G}_m^d$ and $\mathcal{G}_s^d$ to $\mathcal{G}_m^d \cup \mathcal{G}_m^s$ and  $\mathcal{G}_c^d \cup \mathcal{G}_c^s$, respectively. 

\begin{remark}
Theorems \ref{thm: PE-Mean}-\ref{thm: PE} demonstrate that $\delta_{p}$ and $\eta_{p}$ dominate the maximum noise level under the null hypothesis, and select signals under the designated alternatives. As long as $n$ and $p$ are not too small such that $\delta_{p}, \eta_{p} > 1$, which coincides with the high-dimensional framework. 
The theorems confirms the resultant power-enhanced mean test $M_{PE}$ and power-enhanced covariance test $M_{PE}$ satisfy the three PE principles introduced by  \cite{fan2015power}.
\end{remark}


\section{Power-Enhanced Simultaneous Test} \label{subsec: joint-test}

Given two power-enhanced tests, we have boosted the respective power of testing mean vectors and covariance matrices. Before heading to the aggregation of information from the two aspects, we study the joint limiting distribution of the two statistics $M_{PE}$ and $T_{PE}$.

We begin with some insights on the joint distributions for statistics of the two aspects. Suppose we have a random sample i.i.d. drawn from a univariate normal distribution, then it is well-known that the sample mean and sample variance are independent. To a slightly more complex case, suppose we have a random sample i.i.d. drawn from a multivariate normal distribution $N_p(\bmu, \bSigma)$ in the traditional statistical settings when p is fixed. We look into two likelihood ratio test (LRT) statistics. Let $\Lambda_1$ be the LRT statistic for testing $H_0: \{\bmu = \bzero, \bSigma = \bI_p\}$ versus $H_a: \{\bmu \in \mathbb{R}^p , \bSigma = \bI_p\}$, and let  $\Lambda_2$ be the LRT statistic for testing $H_0: \{\bmu \in \mathbb{R}^p, \bSigma = \bI_p\}$ versus $H_a: \{\bmu \in \mathbb{R}^p , \bSigma > \bzero\}$. In multivariate statistics, we know that $\Lambda_1$ and $\Lambda_2$ are independent (Lemma 10.3.1, \cite{anderson2003introduction}).

The above discussion inspires us to conjecture on analogous propositions regarding the joint distribution of $M_{PE}$ and $T_{PE}$. As a matter of fact, in the following theorem, we prove that the two statistics are indeed asymptotically independent. 

\begin{theorem} \label{thm: asymp-dist-H0}
Suppose $n_1/\left(n_1+n_2\right)\rightarrow \gamma$ for some constant $\gamma\in (0,1)$ as $\min\{n_1,n_2\}\rightarrow\infty$ and $\log p = o(n^{1/5})$. Given Assumptions \ref{assum: A2}-\ref{assum: sub-Gaussian}, under $H_0$, as $n_1, n_2 ,p\to \infty$,
\begin{equation}
P\left(M_{PE} \leq x_1, T_{PE} \leq x_2 \right) \rightarrow \Phi(x_1) \Phi(x_2) 
\end{equation}
for any $x_1, x_2 \in \mathbb{R}$.
\end{theorem}

With the information of the two separate power-enhanced tests at hand, the next step is to reasonably aggregate the results for testing means and covariances simultaneously. Most existing works rely on the classical likelihood ratio test \citep{anderson2003introduction} and its variants \citep{jiang2013central,liu2017simultaneous,niu2019lr}. Their test statistics are in the form of a summation of two statistics, where one is designed for detecting discrepancies in covariance matrices, and the other is to catch signals of distinct mean vectors. We call this type of combined statistic as the weighted sum statistics \citep{li2015joint,li2018applications}.

However, the weighted sum statistics bear some drawbacks. The two components are usually of different magnitudes. The combined test would be mostly driven by the statistic with a larger scale, but insensitive to the statistic with a smaller scale. Such inefficiency in combination would lead to power loss in certain alternative spaces. Also, the distribution of the weighted sum statistic depends on the convolution of two marginal distributions, which is usually computationally challenging, resulting in difficulty in choosing 
critical value.

We propose a scale-invariant statistic to simultaneously test the equality of mean vectors and covariance matrices, by combining their separate $p$-values via Fisher's method:
\begin{equation}\label{stat: joint}
J_{n_1,n_2} = -2 \log (p_m) - 2 \log (p_c),
\end{equation}
where $p_m = 1- \Phi\left(M_{PE}\right)$ and  $p_c = 1- \Phi\left(T_{PE}\right)$ are the $p$-values acquired from the power-enhanced mean test and the covariance test, respectively, and $\Phi(\cdot)$ is the cdf of $N(0,1)$. 

As a matter of fact, the Fisher's method has been widely used in meta-analysis for combining the results of multiple scientific studies \citep{hedges2014statistical}. It is worth noticing that meta-analysis is designed for combining studies coming from independent resources. Yet combining two test statistics which are constructed from the same sample would be a different story, and therefore requires careful investigation on the independence assumption.

Theorem \ref{thm: asymp-dist-H0}  proves that under the null hypothesis $H_0$, the two test statistics $M_{PE}$ and  $T_{PE}$ are asymptotically independent. Hence, under $H_0$, $p_m$ and $p_c$ asymptotically independently follow a uniform distribution on the interval $[0,1]$, and therefore $-2\log(p_m)$ and $-2\log(p_c)$ asymptotically independently follow a chi-squared distribution with 2 degrees of freedom. As a result,
\begin{equation}\label{eq: joint-stat-asymp-distribution}
    \text{under }H_{0}: \quad J_{n_1,n_2} \overset{d}{\rightarrow} \chi_4^2 \quad \text{as } n_1,n_2,p\rightarrow\infty.
\end{equation}
Let $q_\alpha$ denote the upper-$\alpha$ quantile of $\chi_4^2$ distribution, we reject the null hypothesis at the significance level $\alpha$ if
\begin{equation}\label{ineq: test-Fisher}
J_{n_1,n_2}\geq q_\alpha.
\end{equation}

\begin{figure}[!bt]
    \centering
    \includegraphics[width=\textwidth]{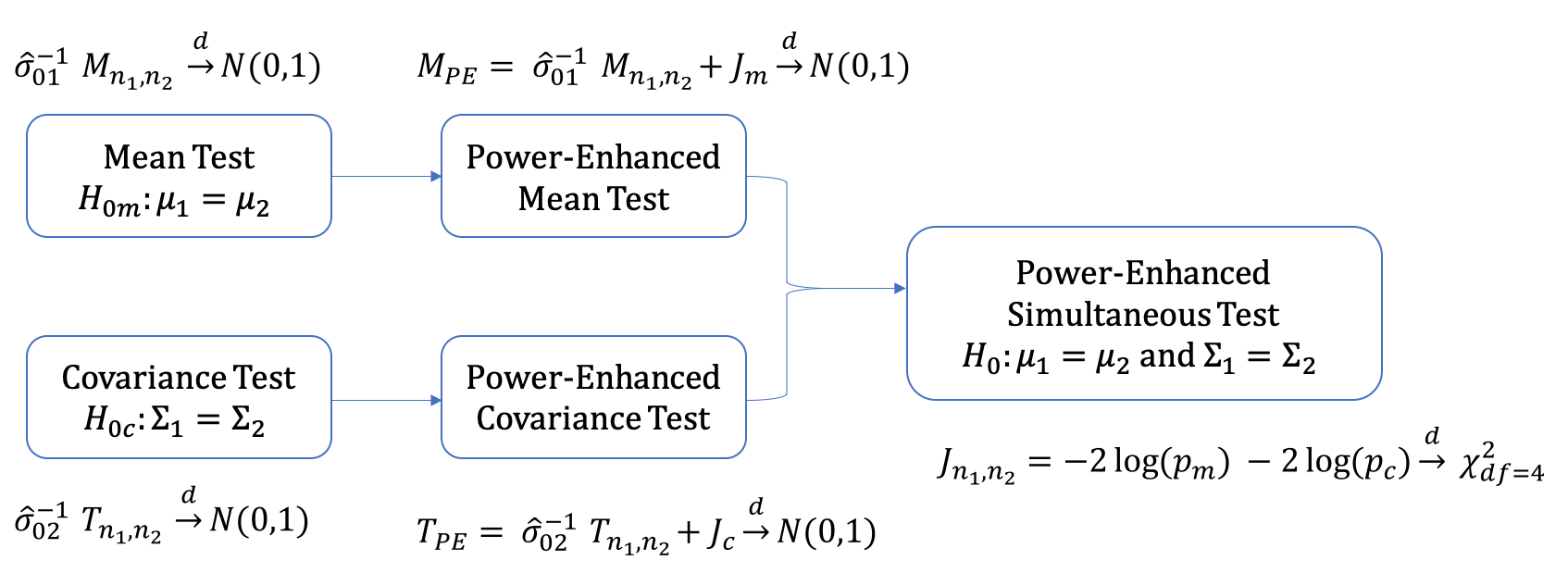}
    \caption{Power-Enhanced Simultaneous Testing Procedure}
    \label{fig: plotFlow}
\end{figure}

The procedures are summarized schematically in Figure \ref{fig: plotFlow}. Equipped with the two key ingredients $M_{PE}$ and $T_{PE}$, we proceed to investigate the size and power property of our proposed test $J_{n_1,n_2}$ in Theorem \ref{thm: simu-test-size}. We show that our proposed test owns asymptotically accurate size approximation to the nominal significance level $\alpha$ and detects differences in either mean vectors or covariances over a wide range of alternatives.

\begin{theorem}[Asymptotic Size and Power for Power-Enhanced Simultaneous Test] \label{thm: simu-test-size}
Suppose $n_1/\left(n_1+n_2\right)\rightarrow \gamma$ for some constant $\gamma\in (0,1)$ as $\min\{n_1,n_2\}\rightarrow\infty$ and $\log p = o(n^{1/5})$. Given Assumptions \ref{assum: A2}-\ref{assum: sub-Gaussian}, as $n_1,n_2,p\rightarrow \infty$, the test $J_{n_1,n_2}$ achieves (i) asymptotically accurate size, that is, under the null hypothesis $H_0:\bmu_1 = \bmu_2 \ \text{  and  } \ \bSigma_1 = \bSigma_2$, we have
    \begin{equation*}
        P\left(J_{n_1,n_2}\geq q_\alpha\right) \rightarrow \alpha,
    \end{equation*}
and (ii) asymptotically consistent power, specifically, 
    $$\inf_{ \{ (\bSigma_1,\bSigma_2)\in \mathcal{G}_c^d \cup \mathcal{G}_c^s \} \cup \{ (\bmu_1, \bmu_2)\in\mathcal{G}_m^d \cup \mathcal{G}_m^s \} } P\left(J_{n_1,n_2}\geq q_\alpha\right) \rightarrow 1.$$
\end{theorem}

There are other ways to aggregate information from the two aspects as the asymptotic independence permits the validity of many other combination methods. In what follows, we present two other tests using different methods to aggregate information to facilitate numerical comparison in the empirical studies. In Section \ref{sec: simulation}, we will show that the Fisher's combined test (\ref{ineq: test-Fisher}) outperforms other approaches as it is asymptotically optimal with respect to Bahadur efficiency.

One is a weighted statistics. Given the asymptotic independence, we may take the sum of squares of two statistics and transform the two asymptotic normal variables to an asymptotic $\chi_2^2$ variable:
\begin{equation}\label{eq: chi2test}
    \text{under }H_{0}: \quad S_{n_1,n_2} = M^2_{PE} + T^2_{PE} \overset{d}{\rightarrow} \chi_2^2 \quad \text{as } n_1,n_2,p\rightarrow\infty.
\end{equation}
The test rejects $H_0$ with a nominal significance level $\alpha$ if $S_{n_1,n_2}\geq c_\alpha$, where $c_\alpha$ is the upper-$\alpha$ quantile of $\chi_2^2$ distribution. 
The other is an alternative p-value combination method. We consider the aggregation via Cauchy transformation \citep{liu2019cauchy}. The Cauchy combination is appealing for its insensitiveness of dependence between the statistics to be combined. In here, even though we obtain the asymptotic independence between $M_{PE}$ and $T_{PE}$, we introduce the Cauchy combination test as a promising alternative. We define the Cauchy combination statistic as follows.
\begin{equation}\label{eq: joint-stat-cauchy-asymp-distribution}
    C_{n_1,n_2} = \frac{1}{2}\tan\left((0.5-p_{m})\pi\right) + \frac{1}{2}\tan\left((0.5-p_{c})\pi\right).
\end{equation}
Under $H_0$, the asymptotic independence ensures that $C_{n_1, n_2}$ converges to a standard Cauchy distribution as $n_1, n_2, p \rightarrow \infty $. The test rejects $H_0$ with a nominal significance level $\alpha$ if $C_{n_1, n_2}\geq k_{\alpha}$, where $k_{\alpha}$ is the upper-$\alpha$ quantile of standard Cauchy distribution.

\section{Simulation Studies} \label{sec: simulation}
In this section, we conduct simulation studies to demonstrate the numerical performance of our proposed power-enhanced simultaneous test. To evaluate the power of the tests under different circumstances, we consider the following three types of alternative hypotheses:
(1) $H_m$: $\bmu_1 \neq \bmu_2,\ \bSigma_1 = \bSigma_2$; (2) $H_c$: $\bmu_1 = \bmu_2,\ \bSigma_1 \neq \bSigma_2$; (3) $H_b$: $\bmu_1 \neq \bmu_2,\  \bSigma_1 \neq \bSigma_2$. 

$H_m$ describes the cases when the two populations share the same covariance matrix but have different means. $H_c$ mimics the opposite situation in which the two populations have the same mean vector but differ in covariances. $H_b$ considers the scenarios that there exist  distinctions in both means and covariances among the two groups. For each alternative, we further consider two types of differences in the parameter of interest: the dense alternatives and the sparse alternatives. We use $H_m^{d}$ and $H_m^{s}$ to represent the existence of dense and sparse differences in $\bmu_1-\bmu_2$, and analogously, $H_c^{d}$ and $H_c^{s}$ to denote those in $\bSigma_1 - \bSigma_2$.

We simulate our samples from the moving average structure shown below, so that we are able to accommodate the complex alternative hypotheses in a general data generating process. For $i=1,\cdots, p,$ let
\begin{align}
\begin{aligned}\label{eq: DGP}
    X_{u,i} & = \mu_{1,i} + Z_{u,i} + \theta_1 Z_{u,i+1}, & u=1,\cdots, n_1, \\
    Y_{v,i} & = \mu_{2,i} + Z_{v+n_1,i} + \theta_2 Z_{v+n_1,i+1}, & v=1,\cdots, n_2, 
\end{aligned}
\end{align}
In such a way, the parameters $\{\mu_{1,i},\mu_{2,i}\}$ alter the mean vectors of our simulated samples $\{\bX_{u}\}_{u=1}^{n_1}$ and  $\{\bY_{v}\}_{v=1}^{n_2}$ to generate $H_m^d$ and $H_m^s$, and $\{\theta_1,\theta_2\}$ control the covariance structure to account for $H_c^d$. By assigning different values to these parameters, we obtain simulated samples with various means and covariances. For the sparse alternatives with respect to covariance matrices $H_c^s$, we generate samples from a different approach by letting $\bX_{u} = \bSigma_{1}^{1/2}\bZ_{u}+\bmu_1$, $\bY_{v} = \bSigma_{2}^{1/2}\bZ_{v+n_1}+\bmu_2$ for $u = 1,\cdots, n_1$, $v = 1, \cdots, n_2$, where $\bZ_{k} = (Z_{k,1}, \cdots, Z_{k,p})'$, $k=1,\cdots, n_1+n_2$.

We first draw $\{Z_{k,i}\}_{1\leq k \leq n_1+n_2, 1\leq i \leq p+1}$ identically and independently from the standard normal $N(0,1)$. To check the robustness to non-normally distributed data, we also generate the random data from the centralized Gamma$(4,2)$. We take the sample sizes as $n_1=n_2=N$ being $100$ and $200$, and let the dimension $p$ take values in $\{100,200,500,800,1000\}$. For each setup, we compare our three proposed testing methods with four existing popular approaches: our proposed power-enhanced simultaneous test $J_{n_1,n_2}$ as in (\ref{ineq: test-Fisher}), the proposed power-enhanced mean test $M_{PE}$ as in (\ref{eq: MPE-stat}), the proposed power-enhanced covariance test $T_{PE}$ as in (\ref{eq: TPE-stat}), the mean test $M_{n_1,n_2}$ proposed by \cite{chen2010two} as in (\ref{stat: mean}), the covariance test $T_{n_1,n_2}$ proposed by \cite{li2012two} as in (\ref{stat: cov}), the $S_{n_1, n_2}$ approximation test as in (\ref{eq: chi2test}), and the Cauchy combination test $C_{n_1,n_2}$ as in (\ref{eq: joint-stat-cauchy-asymp-distribution}). Fore each simulation setting, we report the frequencies of rejections over $5,000$ replications with  $\alpha=0.05$.

To carry out $H_0: \bmu_1 = \bmu_2,\ \bSigma_1 = \bSigma_2$, we set  $\mu_{1i} = \mu_{2i} = 0$ for all $i=1,\cdots, p$, and $\theta_1 = \theta_2 = 0$. Both samples are essentially i.i.d. from $p$-dimensional standard normal or multivariate gamma distribution. To evaluate the power, we fix $\{\mu_{1i}\}_{i=1}^p$ as zeros and $\theta_1 = 0$ for the first population, and vary $\{\mu_{2i}\}_{i=1}^p$ to set up the mean differences in $H_m^d$ and $H_m^s$. As for the covariance alternatives, we change $\theta_2$ to account for dense covariance differences in  $H_c^d$ and implement sparsely differed covariance matrix pair $(\bSigma_1, \bSigma_2)$ to generate $H_c^s$.

As for $H_m$, we set $\theta_2 = 0$ to make sure the two samples share the same covariance matrix. In term of the mean vectors, for $H_m^d$, we follow \cite{benjamini1995controlling} and consider a fixed percentage ($pct$) of violations in $\mu_{1,i} = \mu_{2,i}$ for $i=1,\cdots, p$. 
The nonzero signal strength is determined in a similar fashion to \cite{li2012two} as $\delta = \sqrt{\eta p^{-1/2}}$. To prevent trivial power of $\alpha$ and $1$, we choose $\eta = 0.3$ and $pct = 15\%$. We set $\mu_{2,i} = \delta$ for $1\leq i\leq [p\cdot pct]$ and zeros for the remaining ones. For sparse alternative $H_m^s$, we set the nonzero signal to be $\delta = 0.3 \sqrt{\log p} $ and the number of non-zeros to be $p^r$ with $r = 0.05$.

As for $H_c$, we ensure the two samples share equal means on every dimension. We set $\theta_2 = 0.2$ to create an MA(1) pattern of covariance as the dense alternative $H_c^d$. For the sparse alternative $H_c^s$, we follow \cite{cai2013two} to generate a symmetric sparse matrix $\bU$ with 8 random nonzero entries, each with a magnitude of $\delta = 0.3\sqrt{\log p^2}$. The locations of 4 nonzero entries are randomly selected from the upper triangle of $\bU$ while the other 4 are specified by symmetry. Then we generate samples from $(\bSigma_1, \bSigma_2)$ with $\bSigma_1 = (1+ \varepsilon )\bI_p$ and $\bSigma_2 = (1+ \varepsilon )\bI_p + \bU$, where $\varepsilon = \left|\min\{\lambda_{\min}(\bU+\bI_p), 1\}\right|+0.05$ is to make sure both $\bSigma_1$ and $\bSigma_2$ are positive definite. Finally, with respect to $H_b$, we adopt the same idea as in $H_m$ for the mean differences, and the same approach as in $H_c$ for the covariance differences.

Table \ref{tab: size} presents the empirical size of the seven tests with Normal and Gamma distributed $\{Z_{k,i}\}$ in the data generating process (\ref{eq: DGP}). Tables \ref{tab: Hm-normal}, \ref{tab: Hc-normal}, \ref{tab: Hb-dd-ds-normal} and  \ref{tab: Hb-sd-ss-normal} report the empirical power of the seven methods for testing $H_m$, $H_c$ and $H_b$ with normal distributed $\{Z_{k,i}\}$. We also carry out studies on the power analysis for Gamma distributed $\{Z_{k,i}\}$. The results show a similar 
pattern to the Gaussian cases and are presented in the supplementary material. These numerical comparisons provide us with the following findings:

 \begin{enumerate}
    \item[(1)] Under $H_0$, all of the seven tests achieve reasonably accurate size approximation over a broad range of dimensionality. Besides, the empirical sizes with Gamma distribution illustrate that these tests are quite robust to non-Gaussianity.
    
    \item[(2)] The numerical results of the power-enhanced tests $M_{PE}$ and $T_{PE}$ echo with the power enhancement properties presented in Theorems \ref{thm: PE-Mean} and \ref{thm: PE-Cov}. Table \ref{tab: size} reveals that adding power enhancement components does not inflate the testing size under the null hypothesis $H_0$. On the other hand, Tables \ref{tab: Hm-normal} - \ref{tab: Hb-sd-ss-normal} reflect that the testing power is substantially enhanced under sparse alternatives $H_m^s$ and $H_c^s$.
    
    \item[(3)] As shown in Tables \ref{tab: Hm-normal} and \ref{tab: Hc-normal}, the mean tests ($M_{n_1, n_2}$ and $M_{PE}$) are powerful in detecting mean differences as in $H_m$, but have almost no power in discovering the covariance differences under $H_c$. In contrast, the covariance tests ($T_{n_1, n_2}$ and $T_{PE}$) perform well in declaring significance for covariance alternative $H_c$, however, it is powerless to identify the unequal means under $H_m$.
    
    \item[(4)] With respect to $H_m$ and $H_c$, even though one of the $M_{PE}$ test and $T_{PE}$ test fails, the three combination tests remain powerful across all the experiments. This coincides with the power analysis shown in Theorem \ref{thm: simu-test-size} that the combination of two tests makes use of their respective power under different alternatives, therefore successfully discover the discrepancies in either mean vectors or covariance matrices.
    
    \item[(5)] Tables \ref{tab: Hb-dd-ds-normal} and \ref{tab: Hb-sd-ss-normal} illustrate that our proposed simultaneous test acquires additional gains when both mean differences and covariance differences exist. Under $H_b$, both $M_{PE}$ test and and $T_{PE}$ test successfully sense the differences with regards to the means and covariances respectively. By combining the two tests together, our proposed approach yields to a higher testing power as it can simultaneously detect both types of differences.

    \item[(6)] What's more, for each simulation setting, the proposed test $J_{n_1, n_2}$ prevails with higher power compared with the $S_{n_1, n_2}$ and $C_{n_1, n_2}$ tests. 
    
\end{enumerate}

\begin{onehalfspace}
\begin{table}[H]
\centering
\captionsetup{justification=centering}
\caption{Empirical Size (\%) with Normal and Gamma Distributed $\{Z_{k,i}\}$ \protect\\ in the Data Generating Process}
\label{tab: size}
\begin{tabular}{ccccccc|ccccccc}
\hline
&& \multicolumn{5}{c|}{Normal} & \multicolumn{5}{c}{Gamma} \\
\cline{3-12}
$n$ & Method & $p$ = 100 & 200 & 500 & 800 & 1000 & 100 & 200 & 500 & 800 & 1000 \\
\hline
\multirow{7}{*}{100}
& $M_{n_1, n_2}$     & 5.24 & 5.24 & 5.12 & 5.06 & 5.32 & 5.10 & 4.72 & 5.00 & 5.04 & 5.10 \\ 
& $M_{PE}$  & 5.96 & 5.84 & 5.36 & 5.46 & 5.48 & 5.64 & 5.18 & 5.32 & 5.36 & 5.34 \\ 
& $T_{n_1, n_2}$      & 4.96 & 4.80 & 4.90 & 5.02 & 4.82 & 5.30 & 5.22 & 4.96 & 5.22 & 4.44 \\ 
& $T_{PE}$   & 4.96 & 4.80 & 4.90 & 5.02 & 4.82 & 5.32 & 5.22 & 4.96 & 5.22 & 4.44 \\ 
& $S_{n_1, n_2}$        & 5.70 & 5.84 & 5.98 & 5.12 & 5.40 & 5.92 & 5.60 & 5.34 & 5.10 & 4.84 \\ 
& $C_{n_1, n_2}$      & 5.58 & 5.80 & 6.14 & 5.24 & 5.68 & 5.86 & 5.64 & 5.48 & 5.24 & 5.32 \\ 
& $J_{n_1, n_2}$      & 5.60 & 5.56 & 5.22 & 5.42 & 5.12 & 5.58 & 5.56 & 5.16 & 5.54 & 5.06 \\ 
\hline
\multirow{7}{*}{200}
& $M_{n_1, n_2}$     & 5.48 & 5.30 & 5.46 & 5.16 & 5.22 & 4.94 & 5.06 & 5.06 & 5.24 & 5.26 \\ 
& $M_{PE}$  & 5.68 & 5.56 & 5.62 & 5.18 & 5.30 & 5.34 & 5.32 & 5.18 & 5.32 & 5.34 \\ 
& $T_{n_1, n_2}$      & 4.78 & 4.72 & 5.20 & 4.98 & 4.98 & 4.92 & 5.26 & 4.86 & 5.38 & 5.60 \\ 
& $T_{PE}$   & 4.78 & 4.72 & 5.20 & 4.98 & 4.98 & 4.94 & 5.26 & 4.86 & 5.38 & 5.60 \\ 
& $S_{n_1, n_2}$        & 5.14 & 4.80 & 5.46 & 5.22 & 5.46 & 5.64 & 5.22 & 5.18 & 5.56 & 5.54 \\ 
& $C_{n_1, n_2}$     & 5.36 & 4.86 & 5.22 & 5.36 & 5.30 & 5.30 & 5.00 & 5.30 & 5.60 & 5.72 \\
& $J_{n_1, n_2}$     & 5.50 & 5.12 & 5.24 & 5.30 & 5.08 & 5.50 & 5.22 & 5.56 & 5.54 & 5.54 \\ 
\hline
\end{tabular}

\vspace{1.5ex}
{\small
Note: This table reports the frequencies of rejection by each method under the null hypothesis $H_0$ based on $5000$ independent replications conducted at the significance level $5\%$.}
\end{table}
\end{onehalfspace}

\begin{onehalfspace}
\begin{table}[H]
\centering
\captionsetup{justification=centering}
\caption{Empirical Power (\%) against $H_{m}$ with Normal Distributed $\{Z_{k,i}\}$ \protect\\ in the Data Generating Process} \label{tab: Hm-normal}
\begin{tabular}{c|c|crrrrr}
\hline
$H_m$ & $N$ & Method & \multicolumn{1}{c}{$p$ = 100} & \multicolumn{1}{c}{200} & \multicolumn{1}{c}{500} & \multicolumn{1}{c}{800} & \multicolumn{1}{c}{1000} \\
\hline
\multirow{14}{*}{$H_m^d$} & \multirow{7}{*}{100}
& $M_{n_1, n_2}$     & 47.30 & 44.94 & 47.00 & 46.64 & 46.52 \\
&& $M_{PE}$  & 48.64 & 45.92 & 47.52 & 46.88 & 46.88 \\
&& $T_{n_1, n_2}$      &  5.38 &  5.36 &  4.98 &  5.12 &  4.48 \\
&& $T_{PE}$   &  5.38 &  5.36 &  4.98 &  5.12 &  4.48 \\
&& $S_{n_1, n_2}$        & 34.06 & 30.16 & 30.32 & 29.32 & 28.44 \\
&& $C_{n_1, n_2}$      & 34.22 & 29.88 & 29.96 & 28.88 & 27.98 \\
&& $J_{n_1, n_2}$     & 39.36 & 37.48 & 36.78 & 36.80 & 36.72 \\
\cline{2-8}
& \multirow{7}{*}{200}
& $M_{n_1, n_2}$     & 87.04 & 88.92 & 90.76 & 91.54 & 91.32 \\
&& $M_{PE}$  & 87.34 & 89.10 & 90.86 & 91.58 & 91.34 \\
&& $T_{n_1, n_2}$      &  5.36 &  4.80 &  5.62 &  4.64 &  4.96 \\
&& $T_{PE}$   &  5.36 &  4.80 &  5.62 &  4.64 &  4.96 \\
&& $S_{n_1, n_2}$        & 75.18 & 76.74 & 77.74 & 79.46 & 79.26 \\
&& $C_{n_1, n_2}$      & 75.48 & 77.82 & 78.60 & 80.40 & 79.96 \\
&&$J_{n_1, n_2}$     & 80.32 & 82.20 & 83.32 & 84.46 & 84.64 \\
\hline
\multirow{14}{*}{$H_m^s$} & \multirow{7}{*}{100}
& $M_{n_1, n_2}$     & 42.24 & 33.38 & 54.22 & 44.68 & 17.74 \\
&& $M_{PE}$  & 79.00 & 79.54 & 96.30 & 95.58 & 78.98 \\
&& $T_{n_1, n_2}$      &  4.74 &  4.28 &  4.80 &  4.98 &  5.44 \\
&& $T_{PE}$   &  4.74 &  4.28 &  4.80 &  4.98 &  5.44 \\
&& $S_{n_1, n_2}$        & 76.80 & 78.10 & 95.66 & 95.22 & 77.88 \\
&&$C_{n_1, n_2}$      & 76.92 & 78.04 & 95.60 & 95.20 & 78.02 \\
&& $J_{n_1, n_2}$     & 77.68 & 78.92 & 95.86 & 95.50 & 78.50 \\
\cline{2-8}
& \multirow{7}{*}{200}
& $M_{n_1, n_2}$     & 82.24 & 72.48 &  94.46 & 88.60 & 40.12 \\
&&$M_{PE}$  & 99.22 & 99.68 & 100.00 & 99.98 & 99.74 \\
&& $T_{n_1, n_2}$      &  5.10 &  4.64 &   4.76 &  4.66 &  5.04 \\
&& $T_{PE}$   &  5.10 &  4.64 &   4.76 &  4.66 &  5.04 \\
&& $S_{n_1, n_2}$        & 99.18 & 99.66 & 100.00 & 99.98 & 99.74 \\
&& $C_{n_1, n_2}$     & 99.18 & 99.62 & 100.00 & 99.98 & 99.74 \\
&&$J_{n_1, n_2}$   & 99.18 & 99.64 & 100.00 & 99.98 & 99.74 \\
\hline
\end{tabular}

\vspace{1.5ex}
{\small
Note: (1) This table reports the frequencies of rejection by each method under the alternative hypothesis based on $5000$ independent replications conducted at the significance level $5\%$. \\ \ (2) $H_m$ stands for the type of alternative hypotheses (dense/sparse) in regards of  the mean differences of the two population. }
\end{table}
\end{onehalfspace}

\begin{onehalfspace}
\begin{table}[H]
\centering
\captionsetup{justification=centering}
\caption{Empirical Power (\%) against $H_{c}$ with Normal Distributed $\{Z_{k,i}\}$ \protect\\ in the Data Generating Process } \label{tab: Hc-normal}
\begin{tabular}{c|c|crrrrr}
\hline
$H_c$ & $N$ & Method & \multicolumn{1}{c}{$p$ = 100} & \multicolumn{1}{c}{200} & \multicolumn{1}{c}{500} & \multicolumn{1}{c}{800} & \multicolumn{1}{c}{1000}  \\
\hline
\multirow{14}{*}{$H_c^d$} & \multirow{7}{*}{100}
& $M_{n_1, n_2}$     &  4.80 &  4.92 &  5.12 &  4.84 &  5.10 \\
&& $M_{PE}$  &  5.50 &  5.60 &  5.38 &  5.26 &  5.48 \\
&& $T_{n_1, n_2}$      & 58.62 & 60.38 & 58.90 & 59.12 & 59.98 \\
&& $T_{PE}$   & 58.64 & 60.38 & 58.90 & 59.12 & 59.98 \\
&& $S_{n_1, n_2}$        & 37.22 & 37.52 & 36.12 & 36.52 & 38.16 \\
&& $C_{n_1, n_2}$      & 37.60 & 38.38 & 36.64 & 37.00 & 38.58 \\
&& $J_{n_1, n_2}$     & 45.20 & 46.52 & 46.02 & 45.94 & 47.24 \\
\cline{2-8}
& \multirow{7}{*}{200}
& $M_{n_1, n_2}$     &  5.10 &  5.00 &  5.26 &  5.18 &  5.26 \\
&& $M_{PE}$  &  5.48 &  5.26 &  5.36 &  5.32 &  5.40 \\
&& $T_{n_1, n_2}$      & 97.46 & 98.24 & 98.52 & 98.22 & 98.46 \\
&& $T_{PE}$   & 97.46 & 98.24 & 98.52 & 98.22 & 98.46 \\
&& $S_{n_1, n_2}$        & 91.80 & 92.62 & 93.94 & 92.90 & 93.16 \\
&&$C_{n_1, n_2}$     & 92.62 & 92.98 & 94.24 & 93.76 & 94.08 \\
&& $J_{n_1, n_2}$ & 94.34 & 94.68 & 95.62 & 95.50 & 95.64 \\
\hline
\multirow{14}{*}{$H_c^s$} & \multirow{7}{*}{100}
& $M_{n_1, n_2}$     &  5.12 &  4.98 &  5.14 &  5.28 &  5.22 \\
&& $M_{PE}$  &  5.76 &  5.44 &  5.38 &  5.54 &  5.44 \\
&& $T_{n_1, n_2}$      & 30.94 & 19.48 &  9.78 &  7.66 &  8.14 \\
&& $T_{PE}$   & 65.60 & 66.76 & 58.20 & 45.00 & 37.90 \\
&& $S_{n_1, n_2}$        & 60.00 & 63.74 & 57.06 & 44.02 & 36.72 \\
&&$C_{n_1, n_2}$      & 60.08 & 63.80 & 57.10 & 43.88 & 36.96 \\
&&$J_{n_1, n_2}$    & 62.46 & 65.58 & 57.92 & 44.82 & 37.54 \\
\cline{2-8}
& \multirow{7}{*}{200}
& $M_{n_1, n_2}$     &  5.08 &  5.38 &  5.40 &  4.90 &  5.16 \\
&& $M_{PE}$  &  5.56 &  5.60 &  5.46 &  5.02 &  5.18 \\
&& $T_{n_1, n_2}$      & 72.40 & 46.14 & 17.54 & 11.64 &  9.84 \\
&& $T_{PE}$   & 99.68 & 99.48 & 98.70 & 99.04 & 99.16 \\
&& $S_{n_1, n_2}$        & 99.60 & 99.44 & 98.80 & 98.98 & 99.16 \\
&& $C_{n_1, n_2}$      & 99.58 & 99.44 & 98.82 & 98.98 & 99.16 \\
&& $J_{n_1, n_2}$    & 99.62 & 99.44 & 98.74 & 98.98 & 99.18 \\
\hline
\end{tabular}

\vspace{1.5ex}
{\small
Note: (1) This table reports the frequencies of rejection by each method under the alternative hypothesis based on $5000$ independent replications conducted at the significance level $5\%$. \\ \ (2) $H_c$ stands for the type of alternative hypotheses (dense/sparse) in regards of the covariance differences of the two population. }
\end{table}
\end{onehalfspace}

\begin{onehalfspace}
\begin{table}[H]
\centering
\captionsetup{justification=centering}
\caption{Empirical Power (\%) against $H_{b}$ with Normal Distributed $\{Z_{k,i}\}$ \protect\\ in the Data Generating Process }\label{tab: Hb-dd-ds-normal}
\begin{tabular}{c|c|crrrrr}
\hline
$H_b$ & $N$ & Method & \multicolumn{1}{c}{$p$ = 100} & \multicolumn{1}{c}{200} & \multicolumn{1}{c}{500} & \multicolumn{1}{c}{800} & \multicolumn{1}{c}{1000} \\
\hline
\multirow{14}{*}{$H_{m}^d \cap H_{c}^d $} & \multirow{7}{*}{100}
& $M_{n_1, n_2}$     & 44.78 & 44.80 & 44.40 & 45.26 & 45.60 \\
&& $M_{PE}$  & 46.40 & 45.44 & 44.80 & 45.54 & 45.84 \\
&& $T_{n_1, n_2}$      & 57.80 & 58.70 & 58.94 & 59.16 & 59.78 \\
&& $T_{PE}$   & 57.80 & 58.70 & 58.94 & 59.16 & 59.78 \\
&& $S_{n_1, n_2}$        & 58.80 & 58.12 & 57.58 & 58.08 & 59.30\\
&&$C_{n_1, n_2}$     & 57.22 & 56.56 & 55.40 & 55.76 & 56.72 \\
&& $J_{n_1, n_2}$   & 73.24 & 74.68 & 74.42 & 75.94 & 76.24\\
\cline{2-8}
& \multirow{7}{*}{200}
& $M_{n_1, n_2}$     & 84.44 & 85.58 & 87.92 & 89.20 & 89.22 \\
&& $M_{PE}$  & 84.86 & 85.74 & 88.02 & 89.26 & 89.22 \\
&& $T_{n_1, n_2}$      & 98.16 & 98.26 & 98.26 & 98.38 & 98.48 \\
&& $T_{PE}$   & 98.16 & 98.26 & 98.26 & 98.38 & 98.48 \\
&& $S_{n_1, n_2}$        & 98.84 & 99.12 & 99.38 & 99.24 & 99.30  \\
&&$C_{n_1, n_2}$      & 98.50 & 98.88 & 98.98 & 98.98 & 99.08  \\
&&$J_{n_1, n_2}$  & 99.64 & 99.88 & 99.88 & 99.88 & 99.90 \\
\hline
\multirow{14}{*}{$H_{m}^d \cap H_{c}^s $} & \multirow{7}{*}{100}
& $M_{n_1, n_2}$     & 38.90 & 42.42 & 41.68 & 39.08 & 38.40 \\
&&$M_{PE}$  & 40.52 & 43.12 & 42.16 & 39.42 & 38.70 \\
&& $T_{n_1, n_2}$      & 32.48 & 18.16 &  9.96 &  7.96 &  7.12 \\
&& $T_{PE}$   & 72.36 & 72.62 & 58.12 & 42.76 & 37.58 \\
&& $S_{n_1, n_2}$        & 75.42 & 77.76 & 65.86 & 52.62 & 48.62 \\
&& $C_{n_1, n_2}$      & 75.02 & 77.84 & 65.84 & 52.84 & 48.82 \\
&& $J_{n_1, n_2}$     & 81.34 & 81.98 & 70.56 & 59.42 & 54.56 \\
\cline{2-8}
& \multirow{7}{*}{200}
& $M_{n_1, n_2}$     &  77.06 & 82.92 & 85.12 & 83.38 & 82.46 \\
&&$M_{PE}$  &  77.54 & 83.14 & 85.22 & 83.40 & 82.48 \\
&& $T_{n_1, n_2}$      &  74.24 & 46.52 & 17.68 & 11.54 & 10.06 \\
&& $T_{PE}$   &  99.82 & 99.46 & 98.84 & 98.94 & 99.00 \\
&& $S_{n_1, n_2}$        &  99.90 & 99.72 & 99.40 & 99.38 & 99.34\\
&&$C_{n_1, n_2}$     &  99.90 & 99.72 & 99.42 & 99.36 & 99.34 \\
&&$J_{n_1, n_2}$   &  99.92 & 99.82 & 99.52 & 99.42 & 99.42 \\
\hline
\end{tabular}

\vspace{1.5ex}
{\small
Note: (1) This table reports the frequencies of rejection by each method under the alternative hypothesis based on $5000$ independent replications conducted at the significance level $5\%$. \\ \ (2) $H_b$ stands for the type of alternative hypotheses (dense/sparse) in regards of the mean and covariance differences of the two population. }
\end{table}
\end{onehalfspace}

\begin{onehalfspace}
\begin{table}[H]
\centering
\captionsetup{justification=centering}
\caption{Empirical Power (\%) against $H_{b}$ with Normal Distributed $\{Z_{k,i}\}$ \protect\\ in the Data Generating Process }\label{tab: Hb-sd-ss-normal}
\begin{tabular}{c|c|crrrrr}
\hline
$H_b$ & $N$ & Method & \multicolumn{1}{c}{$p$ = 100} & \multicolumn{1}{c}{200} & \multicolumn{1}{c}{500} & \multicolumn{1}{c}{800} & \multicolumn{1}{c}{1000} \\
\hline
\multirow{14}{*}{$H_{m}^s \cap H_{c}^d $} & \multirow{7}{*}{100}
& $M_{n_1, n_2}$     & 41.16 & 31.58 & 23.28 & 18.20 & 17.70 \\
&& $M_{PE}$  & 77.24 & 77.26 & 77.86 & 77.36 & 78.96 \\
&& $T_{n_1, n_2}$      & 58.18 & 59.96 & 58.80 & 58.26 & 58.78 \\
&& $T_{PE}$   & 58.20 & 59.96 & 58.80 & 58.26 & 58.78 \\
&& $S_{n_1, n_2}$        & 83.14 & 83.90 & 84.86 & 84.50 & 85.36 \\
&& $C_{n_1, n_2}$      & 83.14 & 84.06 & 84.98 & 84.64 & 85.72 \\
&& $J_{n_1, n_2}$    & 87.50 & 88.68 & 88.60 & 88.00 & 88.80 \\
\cline{2-8}
& \multirow{7}{*}{200}
& $M_{n_1, n_2}$     & 81.52 &  69.82 &  51.78 &  41.72 & 38.26 \\
&& $M_{PE}$  & 99.10 &  99.46 &  99.60 &  99.74 & 99.74 \\
&& $T_{n_1, n_2}$      & 97.56 &  97.78 &  98.40 &  97.96 & 98.54 \\
&& $T_{PE}$   & 97.56 &  97.78 &  98.40 &  97.96 & 98.54 \\
&& $S_{n_1, n_2}$        & 99.96 &  99.96 & 100.00 &  99.98 & 99.98 \\
&& $C_{n_1, n_2}$      & 99.96 &  99.98 & 100.00 & 100.00 & 99.98 \\
&&$J_{n_1, n_2}$     & 99.98 & 100.00 & 100.00 & 100.00 & 99.98 \\
\hline
\multirow{14}{*}{$H_{m}^s \cap H_{c}^s $} & \multirow{7}{*}{100}
& $M_{n_1, n_2}$     & 34.84 & 29.34 & 20.34 & 16.94 & 15.04 \\
&& $M_{PE}$  & 66.12 & 71.62 & 70.74 & 67.76 & 68.60 \\
&& $T_{n_1, n_2}$      & 30.74 & 19.70 &  9.84 &  7.76 &  7.12 \\
&& $T_{PE}$   & 69.44 & 74.36 & 57.28 & 43.96 & 38.32 \\
&& $S_{n_1, n_2}$        & 85.56 & 89.86 & 84.80 & 78.64 & 78.14 \\
&& $C_{n_1, n_2}$     & 85.48 & 89.90 & 84.84 & 78.90 & 78.10 \\
&& $J_{n_1, n_2}$   & 87.30 & 91.14 & 86.12 & 79.82 & 79.44 \\
\cline{2-8}
& \multirow{7}{*}{200}
& $M_{n_1, n_2}$     & 72.94 & 64.52 & 46.32 & 36.32 & 31.20 \\
&& $M_{PE}$  & 97.54 & 98.48 & 98.92 & 98.78 & 98.68 \\
&& $T_{n_1, n_2}$      & 74.22 & 46.46 & 17.76 & 11.72 & 10.94 \\
&& $T_{PE}$   & 99.72 & 99.36 & 99.00 & 98.86 & 98.88 \\
&& $S_{n_1, n_2}$        & 99.98 & 99.94 & 99.94 & 99.80 & 99.76 \\
&& $C_{n_1, n_2}$      & 99.98 & 99.94 & 99.94 & 99.84 & 99.76 \\
&& $J_{n_1, n_2}$    & 99.98 & 99.94 & 99.96 & 99.84 & 99.76 \\
\hline
\end{tabular}

\vspace{1.5ex}
{\small
Note: (1) This table reports the frequencies of rejection by each method under the alternative hypothesis based on $5000$ independent replications conducted at the significance level $5\%$. \\ \ (2) $H_b$ stands for the type of alternative hypotheses (dense/sparse) in regards of the mean and covariance  differences of the two population. }
\end{table}

\end{onehalfspace}

\begin{figure}[H]
    \centering
    \includegraphics[width=0.8\textwidth, height = 3in]{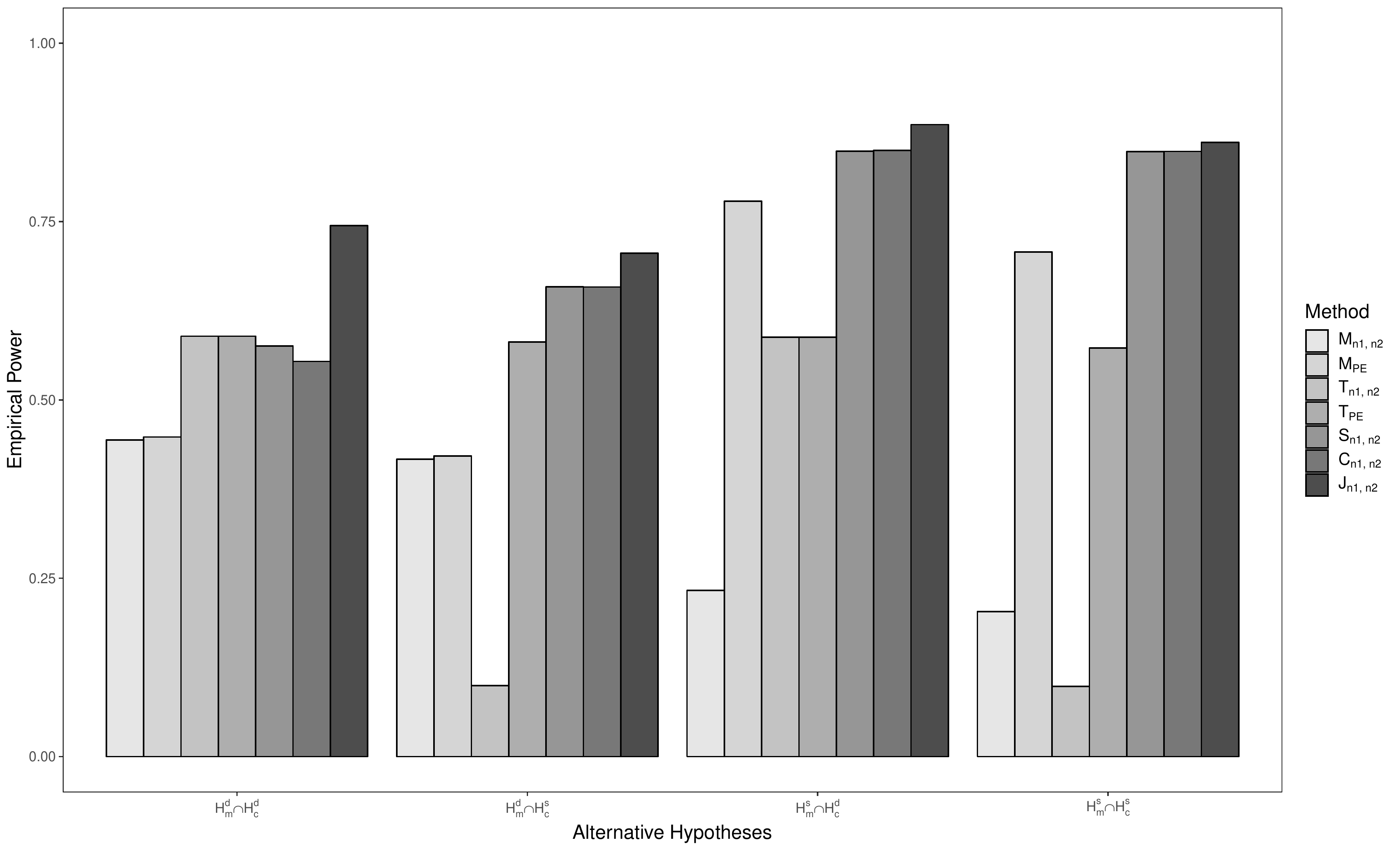}
    \captionsetup{justification=centering}
    \caption{Empirical Power Comparison of the Seven Tests Under $H_b$  \protect\\ with Normal Distributed $\{Z_{k,i}\}$ and $N = 100$, $p = 500$ }
    \label{fig: plotHb}
\end{figure}

In addition, 
Figure \ref{fig: plotHb} provides a graphical representation of the testing power using seven approaches under $H_b$ when both mean differences and covariances differences exist. We study four different hypotheses consisting of the combination of sparsely/densely differed means ($H_m^s/H_m^d$) and sparsely/densely differed covariances ($H_c^s/H_c^d$). The figure shows that the tests $M_{n_1, n_2}$ and $T_{n_1, n_2}$ favor dense alternatives $H_m^d$ and $H_c^d$, respectively, because of its nature of quadratic forms, but lack the ability of detecting sparse alternatives such as $H_m^s$ and $H_c^s$. Fortunately, the proposed power-enhanced tests $M_{PE}$ and $T_{PE}$ greatly promote the respective testing power under $H_m^s$ and $H_c^s$. When it comes to jointly testing means and covariances, the plot clearly shows that our proposed Fisher's combined test $J_{n_1,n_2}$ achieves the highest power among the three combination approach ($F_{n_1, n_2}$, $S_{n_1, n_2}$ and $C_{n_1, n_2}$).

In summary, the simulation results demonstrate the promising finite-sample performance of our proposed simultaneously tests, providing numerical evidence to verify the theoretical properties introduced in the previous section. Under the null hypothesis, the proposed test retains the desired nominal significance level. It is powerful in detecting either mean differences or covariance differences, and it remains high power for either sparse alternatives or dense alternatives. Moreover, the testing power is boosted against more general alternatives.

\section{Application to Gene-Set Testing}\label{sec: realdata}

This section demonstrates the power of our proposed tests through a real application on an Acute Lymphoblastic Leukemia (ALL) dataset from the Ritz Laboratory at the Dana-Farber Cancer Institute (DFCI). The data was originally published by \cite{chiaretti2004gene} and is now available at the \href{https://www.bioconductor.org/}{\color{blue}{Bioconductor}} website. The ALL dataset contains gene expression levels of 12,625 probes on Affymetrix chip series HG-U95Av2 from 128 individuals with either T-cell ALL or B-cell ALL, depending on the type of lymphocyte for the leukemia cells. This study focuses on a subset of the ALL data for 79 patients with the B-cell ALL. We further divide the patients into two groups according to their B-cell tumors' subtypes: the BCR/ABL fusion and the cytogenetically normal NEG, whose sample sizes are 37 and 42, respectively.

Identifying differentially expressed gene-sets has received considerable attention in genetic studies \citep{efron2007testing,goeman2007analyzing}. Since each gene does not work individually but rather tend to function groups to achieve complex biological tasks, researchers look into gene expression profiles based on groups of genes depending on their functional characteristics. To make full use of prior biological knowledge, we group sets of genes according to their Gene Ontology (GO) annotations. The GO system describes the biological domains with respect to three aspects: biological process (BP), cellular component (CC), and molecular function (MF). We follow the same criteria to perform a pre-screening procedure  by excluding those probes with low fluorescence intensities and narrowly spread, characterized by small absolute values and small interquantile ranges. The filtering step retains 2,391 probes, corresponding to 1849 unique GO terms in BP category, 306 in CC and 324 in MF. 

\begin{figure}[H]
\centering
\includegraphics[width = 0.8\textwidth]{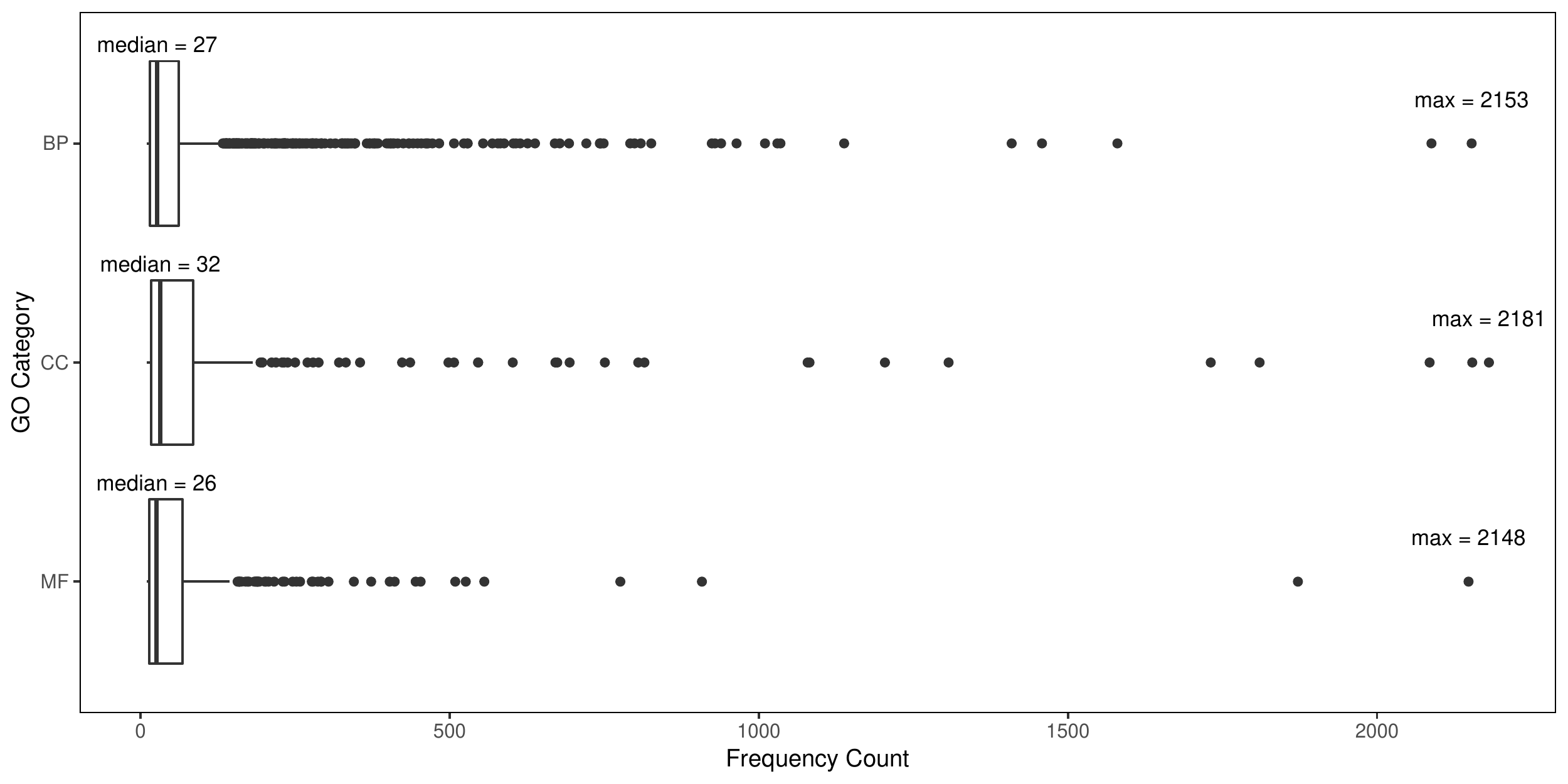}
\caption{Boxplots of the Dimension of Gene-sets for Three GO Categories}\label{fig: summary-gene-set}
\end{figure}

Let $S_1,\cdots, S_K$ denote $K$ gene-sets and  $\{\bmu_{1S_k}, \bSigma_{1S_k}\}$, $\{\bmu_{2S_k}, \bSigma_{2S_k}\}$ be the mean vectors and covariance matrices of two types of tumors respectively. We are interested in testing 
$$H_{0,category}: \bmu_{1S_k} = \bmu_{2S_k}\ \text{ and }\ \bSigma_{1S_k} = \bSigma_{2S_k},\quad k=1,\cdots,K$$
where $category \in \{BP, CC, MF\}$. We classify gene-sets into three different GO categories and shall test each GO category separately. Figure \ref{fig: summary-gene-set} plots the dimension of gene-sets contained in each category. The dimension of gene-sets in each category can be as large as two thousand, which is much larger than the sample sizes $n_1=37$ and $n_2=42$. 

\begin{figure}[H]
\centering
\includegraphics[width=5in, height=2.7in]{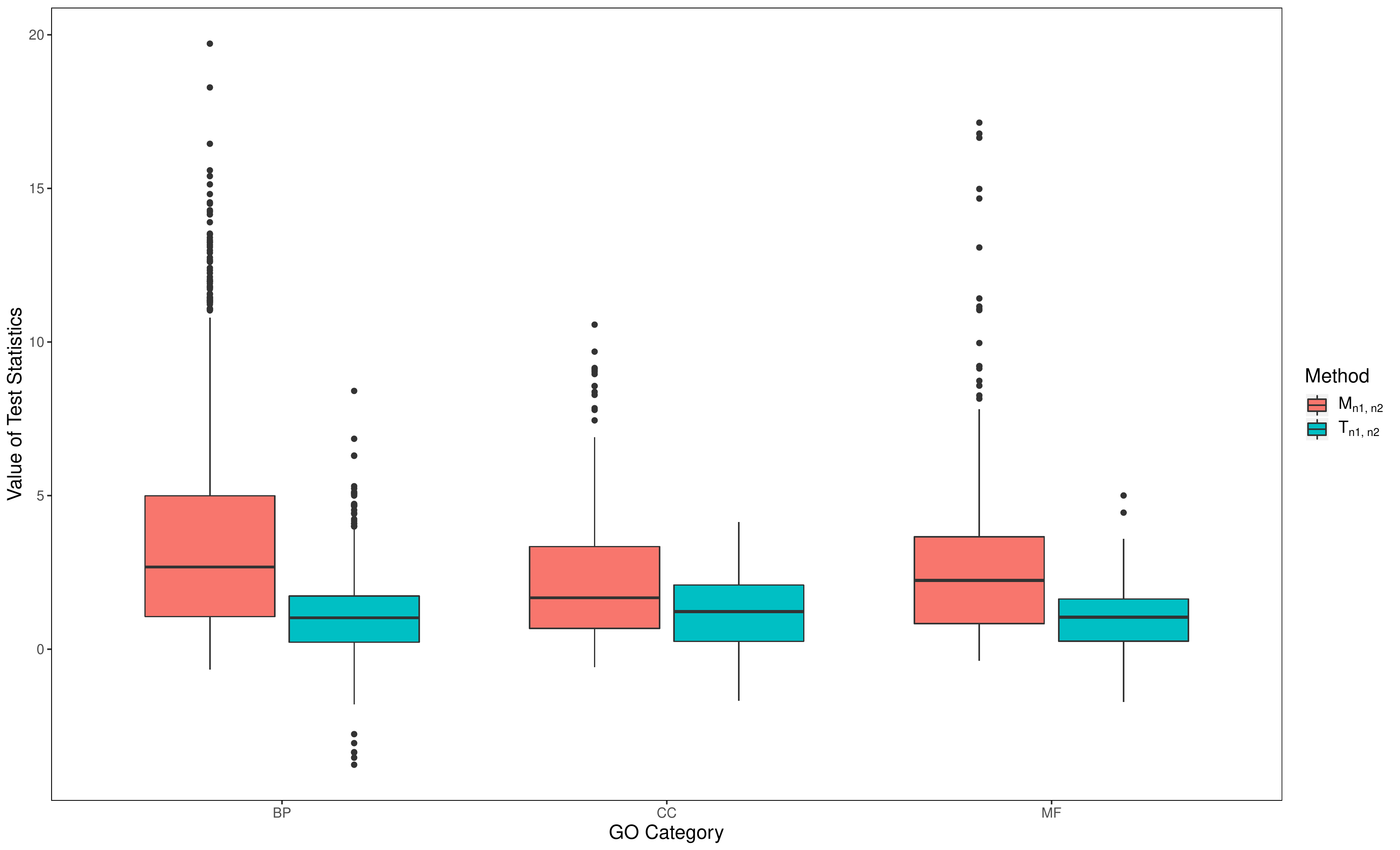}
\caption{Boxplots of the $\widehat{\sigma}_{01}^{-1}M_{n_1, n_2}$ and $\widehat{\sigma}_{02}^{-1}T_{n_1, n_2}$ Test Statistics for Three GO Categories}\label{fig: plotstat}
\end{figure}

Before proceeding, we explore the compuated values of power-enhanced test statistics $M_{n_1, n_2}$ and $T_{n_1, n_2}$ for all gene-sets. Figure \ref{fig: plotstat} presents boxplots of  $M_{n_1,n_2}$ and $T_{n_1,n_2}$ within each GO category.  The $M_{n_1, n_2}$ statistics have relatively larger values compared with the $T_{n_1, n_2}$ statistics. Recall that under the null hypothesis, both statistics converge to $N(0,1)$ in distribution. The finding that the $M_{n_1, n_2}$ statistics have larger absolute values indicates that for these gene-sets, their mean vectors are more different compared to the covariance matrices between the two groups. Moreover, considering significance level $\alpha=0.05$ and the upper $\alpha$-quantile of $N(0,1)$ $z_\alpha = 1.645$, a large number of $M_{n_1, n_2}$ statistics fall above the threshold $z_\alpha$. Therefore we would expect a lot of rejections for testing the equality of the mean vectors. The discussions in this paragraph give us an exploratory view of the dataset. Later on, we will present more precise comparisons among various test approaches.

We then apply our power-enhanced simultaneous test $J_{n_1,n_2}$ to test the means and covariances simultaneously, together with the mean test $M_{n_1, n_2}$, the covariance test $T_{n_1, n_2}$ and the two power-enhanced tests $M_{PE}$ and $T_{PE}$. We compare our proposed approach $J_{n_1,n_2}$ with the $\chi^2$ approximation $S_{n_1, n_2}$ as well as the Cauchy combination $C_{n_1, n_2}$. 
In order to control the false discovery rate (FDR), we apply the Benjamini-Hochberg (BH) procedure \citep{benjamini1995controlling} to each GO category. Table \ref{tab: realdataBH} reports the number of significant gene-sets declared by different tests with nominal level $\alpha=0.05$ for every category.

\begin{table}[H]
\centering
\small
\captionsetup{justification=centering}
\caption{The Number of Significant Gene-sets Declared by Different Tests \protect \\ after BH Control with Nominal Level $\alpha=0.05$} \label{tab: realdataBH}
\begin{tabular}{c|c|C{1.5cm}|C{1.5cm}|C{1.5cm}}
\hline
\multicolumn{2}{c|}{GO Category} & \multicolumn{1}{c|}{BP} & \multicolumn{1}{c|}{CC} & \multicolumn{1}{c}{MF} \\
\hline
\multicolumn{2}{c|}{Total number of Gene-sets} &1849 & 306 & 324 \\
\hline
\multirowcell{7}{Number of \\ Significant \\ Gene-sets}
& $M_{n_1, n_2}$ &  1134 & 140 & 183  \\
& $M_{PE}$ & 1469 & 216 & 236 \\
& $T_{n_1, n_2}$ &  126 & 55 & 20\\
& $T_{PE}$ & 126 & 55 & 20 \\
& $S_{n_1, n_2}$ & 1485 & 219 & 234 \\
& $C_{n_1, n_2}$ & 1484 & 220 & 233\\
& $J_{n_1, n_2}$ &  1511 & 226 & 238 \\
\hline
\end{tabular}
\end{table}

As shown in Table \ref{tab: realdataBH}, $J_{n_1,n_2}$ identifies more significant gene-sets than the other methods. The $M_{n_1, n_2}$ test declares a lot of significance whereas the $T_{n_1, n_2}$ test only identifies a few. The $M_{PE}$ identifies a few more differentially expressed gene-sets with respect to mean vectors, while the $T_{PE}$ does not yield additional power in detecting the differences among covariances. This indicates there exist a large number of unequal means between the two types of tumors, but not much differences in their covariance patterns. This phenomenon emphasizes the importance of developing a powerful method for jointly testing the means and covariances, so that we have a better chance to detect differences between two distributions even though we are in lack of prior knowledge about whether the differences reside in means or covariances.

The $\chi^2$ approximation $S_{n_1, n_2}$ and the Cauchy combination $C_{n_1, n_2}$ yield comparative performance. As shown in Table \ref{tab: realdataBH}, the two methods identify more differences than the covariance test $T_{PE}$, yet potentially miss some differentially expressed gene-sets compared to the mean test $M_{PE}$. In contrast, our proposed $J_{n_1,n_2}$ is able to identify more discrepancies between the two groups, compared to the other three combination approaches and also compared to the original means test as well as the covariance tests. In a short summary, our proposed Fisher's combined simultaneous test $J_{n_1,n_2}$ benefits from incorporating the information from the mean tests and covariance tests, and outperforms other combination methods in detecting the significant differences among the gene-sets.

Next, we study those gene-sets which are declared significant only by $J_{n_1,n_2}$ but not any other method. Among those, we pay special attention to the GO-term ``GO:0005125" in the MF category. This gene-set contributes to cytokine activity, including interleukins which are a group of cytokines that regulate inflammatory and immune responses \citep{okada2004cytokine}. Extensive scientific studies have revealed the close relationships between interleukins and leukemia \citep{touw1990interleukin,paietta1997expression,yoda2010functional,canale2011interleukin}. For another example, it is known microRNAs act complementarily to regulate disease-related mRNA modules in human diseases \citep{chavali2013micrornas}. We observe that the expression levels of ``GO:0006913" in the BP category are statistically different between the two groups. This GO-term refers to nucleocytoplasmic transport, whose association with leukemia has been validated by numerous cancer studies \citep{chavali2013micrornas,gravina2014nucleo, takeda2014nucleoporins}.
These biological evidences suggest our power-enhanced simultaneous test $J_{n_1,n_2}$ provides more useful information compared with other approaches, which further implies the importance of developing power-enhanced simultaneous tests.

\section{Conclusion} \label{sec: conclusion}

In this work, we study the problem of jointly testing the equality of two-sample mean vectors and covariance matrices of high-dimensional data. We introduce a new power-enhanced simultaneous test, and prove the test achieves accurate asymptotic size, enhanced and consistent asymptotic power under a more general alternative, and asymptotic optimality with respect to Bahadur efficiency. The proposed test is scale-invariant and computationally efficient. We demonstrate the finite-sample performance using simulation studies and a real application to gene-set testing.

{
\bibliographystyle{agsm}
\bibliography{ref-test-meancov}
}

\end{document}